\documentclass[showkeys,showpacs,amsmath,amssymb,a4paper,12pt]{revtex4}
\usepackage{amsmath}
\usepackage{amssymb}
\usepackage{amsmath,amsthm,amssymb,amscd}
\usepackage{latexsym}
\usepackage{indentfirst}
\usepackage{subfigure}
\usepackage{graphicx}
\usepackage[top=1in,bottom=1in,left=1.25in,right=1.25in]{geometry}

\makeatother
\theoremstyle{plain}

\begin{document}

\title{
  \bf Shock of three-state model for intracellular transport of  kinesin KIF1A
}

\author{Yunxin Zhang}\email[Email: ]{xyz@fudan.edu.cn}
\affiliation{Shanghai Key Laboratory for Contemporary Applied Mathematics,
Centre for Computational System Biology,\\
School of Mathematical Sciences, Fudan University, Shanghai 200433, China.
}

\begin{abstract}
Recently, a three-state model is presented to describe the intracellular traffic of unconventional (single-headed) kinesin KIF1A [Phys. Rev. Lett. {\bf 95}, 118101 (2005)], in which each motor can bind strongly or weakly to its microtubule track, and each binding site of the track might be empty or occupied by one motor. As the usual two-state model, i.e. the totally asymmetric simple exclusion process (TASEP) with motor detachment and attachment, in steady state of the system, this three-state model also exhibits shock (or domain wall separating the high-density and low density phases) and boundary layers. In this study, using mean-field analysis, the conditions of existence of shock and boundary layers are obtained theoretically. Combined with numerical calculations, the properties of shock are also studied. This study will be helpful to understand the biophysical properties of the collective transport of kinesin KIF1A.
\end{abstract}




\maketitle \baselineskip=6mm

\baselineskip=22pt

\section{Introduction}
Molecular motors are biogenic force generators acting in the
nanometer range. They are responsible for
intracellular transport of wide varieties of cargo from one
location to another in eukaryotic cells \cite{Howard2001, Bray2001, Cooper2000, Schliwa2003, Sperry2007, Vale2003}. Linear motors produce
sliding movements along filamentous structures called protein
tracks; for example, myosin slides along actin
filament \cite{Mehta1999, Anneka2007, Christof2006, Katsuyuki2007}, kinesin \cite{Block1990, Vale1996, Yildiz2004, Carter2005, Guydosh2009} and dynein
\cite{Mallik2004, Samara2006, Toba2006, Gennerich2007, Houdusse2009} along  microtubule.
Microtubule and filamentary protein actin are protein filaments
which form part of a dual-purpose scaffolding called cytoskeleton,
act like struts or girders for the cellular architecture and, at
the same time, also serve as tracks for the intracellular
transportation networks.

However, experiments found, one single filamentary track is usually traveled along by multiple motors. Fundamental understanding of these collective
physical phenomena may expose the causes of motor-related
diseases (e.g., Alzheimer's disease). In literatures, these phenomena are usually described by the totally asymmetric simple exclusion process (TASEP), which is
originally proposed in \cite{MacDonald1968}, consists of
particles hopping unidirectionally with hard-core exclusion along a
1D lattice. TASEP is one of many examples for driven
systems with stationary nonequilibrium states, which cannot be
described in terms of Boltzmann weights.
To model the attachment and detachment of motors to and from tracks, Parmeggiani
{\it et al} \cite{Parmeggiani2003, Parmeggiani2004} discussed
a class of driven lattice gas obtained by coupling 1D TASEP to
Langmuir kinetics, in which the attachment and detachment of motors is modeled as particle creation and annihilation
respectively. Furthermore, Lipowsky {\it et al}
\cite{Lipowsky2001, Lipowsky2006} suggested a more general
model, in which the diffusion of motors in the cell is considered
explicitly. However, in reality, a motor protein is not a mere
particle, but an enzyme whose mechanical movement is coupled with
its biochemical cycle. Therefore, recently, a three-state model is presented by Nishinari {\it et al} to describe the intracellular transport of single-headed kinesin KIF1A \cite{Nishinari2005, Nishinari2007}. In which, the microtubule (MT) binding motor might be in two states: strongly MT binding state and weakly MT binding state, denoted by $S, W$ respectively. Biochemically, the strongly binding state corresponds to bare motor or ATP binding motor state, and the weakly binding state corresponds to ADP binding state.

One of the important feature of the collective motion of motors along one single track is the possible appearance of shock or domain wall, which is
defined as the interface between the low-density and high density
regions. For the usual two-state TASEP, using mean field method, the existence and properties of the shock have been discussed recently \cite{Zhang2007}. In this study, similar analysis to the Nishinari's three-state model will be presented, the
efficient and necessary conditions of the existence of shock will be given theoretically, and with the aid of numerical calculations the properties of the shock will also be discussed. The method used in this study can be regarded as a generalization of the one presented in \cite{Zhang2007}.

In the next section, the three-state model and its mean field approximation will be briefly introduced, and then in Sec. III the conditions of the existence of shock will be presented. The existence of boundary layers and the properties of shock will be discussed in Sec. IV and V. Finally, this study will be shortly summarized in Sec. VI.

\section{TASEP with three internal states}
The three-state TASEP given in \cite{Nishinari2007} can be mathematically desccribed as follows. Let $S_i$ and $W_i$ denote the probabilities of finding a molecular motor in the states 1 and 2 at the lattice site $i$ at time $t$ respectively (states 1 and 2 correspond to strongly bound and weakly bound states of molecular motors). Then $S_i, W_i$ are governed by the following master equations
\begin{equation}\label{eq1}
\begin{aligned}
\frac{dS_i}{dt}=&\omega_a(1-S_i-W_i)-\omega_h S_i-\omega_d
S_i+\omega_sW_i\cr &+\omega_f
W_{i-1}(1-S_i-W_i)+(1-c)\omega_fW_i(S_{i+1}+W_{i+1}),
\end{aligned}
\end{equation}
\begin{equation}\label{eq2}
\begin{aligned}
\frac{dW_i}{dt}=&-(\omega_s+\omega_f)W_i(1-S_{i+1}-W_{i+1})+\omega_h
S_i\cr &-[\omega_s+(1-c)\omega_f]W_i(S_{i+1}+W_{i+1})\cr
&-\omega_b W_i(2-S_{i+1}-W_{i+1}-S_{i-1}-W_{i-1})\cr
&+\omega_b(W_{i-1}+W_{i+1})(1-S_i-W_i),
\end{aligned}
\end{equation}
where $\omega_a$ is the rate of a molecular motor binding to the
empty lattice site $i$, i.e. the transition rate of state 0 to state
1, $\omega_h$ is the transition rate of state 1 to state 2, i.e. the
rate of ATP hydrolysis, $\omega_d$ is the transition rate of state 1
to state 0, i.e. the rate of detachment, $\omega_b$ is the rate of
random Brownian motion. After the release of ADP, the motor steps
forward to the next binding site in front with rate $\omega_f$,
stays at the current location with rate $\omega_s$. $c$ is an
interpolating parameter ($0\le c\le 1$). The corresponding equations
for the left boundary ($i=1$) are given by
\begin{equation}\label{eq3}
\begin{aligned}
\frac{dS_1}{dt}=&\alpha(1-S_1-W_1)-\omega_h S_1-\gamma_1
S_1+\omega_sW_1+(1-c)\omega_fW_1(S_{2}+W_{2}),
\end{aligned}
\end{equation}
\begin{equation}\label{eq4}
\begin{aligned}
\frac{dW_1}{dt}=&-(\omega_s+\omega_f)W_1+\omega_h S_1+c\omega_f
W_1(S_{2}+W_{2})\cr &+\omega_b W_2(1-S_{1}-W_{1})-\omega_b
W_{1}(1-S_2-W_2)-\gamma_2 W_1,
\end{aligned}
\end{equation}
where $\alpha$ is the rate of attachment of the motors at the left
boundary (i.e. the lattice site $i=1$), $\gamma_1$, $\gamma_2$ are
the rates of detachment of motors in state 1 and state 2 at the left
boundary respectively. The equations for the right
boundary ($i=N$)  are give by
\begin{equation}\label{eq5}
\begin{aligned}
\frac{dS_N}{dt}=&\delta(1-S_N-W_N)+\omega_fW_{N-1}(1-S_N-W_N)\cr
&+\omega_s W_N-\omega_h S_N-\beta_1 S_N,
\end{aligned}
\end{equation}
\begin{equation}\label{eq6}
\begin{aligned}
\frac{dW_N}{dt}=&\omega_h S_N-\omega_s W_N-\beta_2 W_N+\omega_b
W_{N-1}(1-S_N-W_L)\cr &-\omega_b W_N(1-S_{N-1}-W_{N-1}),
\end{aligned}
\end{equation}
where $\delta$ is the rate of attachment of the motors at the
right boundary (i.e. the lattice site $i=N$), $\beta_1$ $\beta_2$
are the rates of detachment of motors in state 1 and state 2 at
the right boundary respectively.

It should be pointed out that the exclusion process described above
is different from the one discussed in \cite{Reichenbach2006}, where multiple
occupancy of sites is allowed if particles are in different internal
states. Here, the multiple occupancy is unallowed. However, the
particles bounding to the lattice site might be in two different
states 1 and 2, corresponding to the strongly bound and weakly bound
states. As in \cite{Parmeggiani2003}, attachment and detachment of a
motor are modeled as, effectively, creation and annihilation of
motors on the lattice. Moreover, the transition between states 1
and 2 is described by the rates $\omega_h,\ \omega_s,\ \omega_f$,
the Brownian ratched mechanism is described by rate $\omega_b$.

{\bf Mean Field Approximation: } In the large $N$ limit, we can make the continuum mean field
approximation to Eqs. (\ref{eq1}) and (\ref{eq2}). Let $\Delta
x=\frac{1}{N-1}$ and $x=(i-1)\triangle x$. Obviously, $0\le x\le 1$,
since $1\le i\le N$. Using the Taylor expansion
\begin{equation}\label{eq7}
S(x\pm\Delta x)=S(x)+\sum_{k=1}^{+\infty}\frac{(\pm\Delta
x)^k}{k!}\frac{\partial^kS(x)}{\partial x^k},\quad W(x\pm\Delta
x)=W(x)+\sum_{k=1}^{+\infty}\frac{(\pm\Delta
x)^k}{k!}\frac{\partial^kW(x)}{\partial x^k}.
\end{equation}
The continuum limits of Eqs. (\ref{eq1}) and (\ref{eq2}) are then
\begin{equation}\label{eq8}
\begin{aligned}
\frac{\partial S(x,t)}{\partial t}=&\omega_a(1-S-W)+\omega_s
W-(\omega_h+\omega_d)S\cr &+\omega_f\left( W-\Delta x\frac{\partial
W(x,t)}{\partial x}\right)(1-S-W)\cr & +(1-c)\omega_f
W\left(S+W+\Delta x\frac{\partial S(x,t)}{\partial x}+\Delta
x\frac{\partial W(x,t)}{\partial x}\right)\cr &+O(\Delta x^2),
\end{aligned}
\end{equation}
\begin{equation}\label{eq9}
\begin{aligned}
\frac{\partial W(x,t)}{\partial t}=&c\omega_f W\left(S+W+\Delta
x\frac{\partial S(x,t)}{\partial x}+\Delta x\frac{\partial
W(x,t)}{\partial x}\right)\cr &-(\omega_s+\omega_f)W+\omega_h
S+O(\Delta x^2).
\end{aligned}
\end{equation}
Thus, the probability density $\rho(x,t)=S(x,t)+W(x,t)$ of finding
a molecular motor at lattice site $x$ at time $t$ satisfies [summing Eqs. (\ref{eq8}) and (\ref{eq9})]
\begin{equation}\label{eq10}
\begin{aligned}
\frac{\partial \rho(x,t)}{\partial t}=&\omega_a(1-\rho)-\omega_d
S+\omega_f\frac{\partial W(\rho-1)}{\partial x}\Delta x+O(\Delta
x^2).
\end{aligned}
\end{equation}

As the discussion in \cite{Popkov2003}, in the thermodynamic limit
$N\to \infty$, there are three regimes to be distinguished. If
$\omega_a$ and $\omega_d$ are of order $[1/(N-1)]^\alpha$ with $\alpha<1$, then at
the steady state, the system, Eqs. (\ref{eq9}) and (\ref{eq10}), reduces to
\begin{equation}\label{eq11}
\left\{\begin{aligned} &\omega_a(1-W-S)-\omega_d S=0,\cr &c\omega_f
W\left(S+W\right)-(\omega_s+\omega_f)W+\omega_h S=0.
\end{aligned}\right.
\end{equation}
So the probability $S$ satisfies
\begin{equation}\label{eq12}
ck(k+1)S^2+[(k+1)(l+1)+n-c(2k+1)]S+(c-l-1)=0,
\end{equation}
and $W=1-(k+1)S$, $\rho=S+W$, where $l=\omega_s/\omega_f,\
n=\omega_h/\omega_f,\ k=\omega_d/\omega_a$ (because of the
particle-hole symmetry, we restrict the discussion to the case
$\omega_a\ge \omega_d$, i.e. $0\le k\le 1$).

For the cases that $\omega_a$ and $\omega_d$ are of order $[1/(N-1)]^\alpha$ but with $\alpha>1$, the local
kinetics is negligible and the system will be
\begin{equation}\label{eq13}
\left\{\begin{aligned} &W(W+S-1)=C,\cr &c\omega_f
W\left(S+W\right)-(\omega_s+\omega_f)W+\omega_h S=0
\end{aligned},\right.
\end{equation}
where the constant $C$ is determined by the left or right boundary conditions \cite{Derrida1993, Schutz1993}. The case of the local rates $\omega_a$ and $\omega_d$ being of the order $1/(N-1)$ is the most interesting one, and will
be investigated further in this study. In the following, we
always assume that the local rates $\omega_a$ and $\omega_d$ are of
the order $1/(N-1)$.

\section{The existence of shock}
Let
\begin{equation}\label{eq14}
\Omega_a=\frac{\omega_a}{\Delta
x},\qquad\Omega_d=\frac{\omega_d}{\Delta x},
\end{equation}
then at steady state, the leading terms of $\Delta x$ of Eqs. (\ref{eq9}) (\ref{eq10}) are
\begin{equation}\label{eq15}
\left\{\begin{aligned} &\omega_f\frac{\partial W(\rho-1)}{\partial
x}+\Omega_a(1-\rho)-\Omega_d S=0\cr &c\omega_f
W\left(S+W\right)-(\omega_s+\omega_f)W+\omega_h S=0
\end{aligned}\right.
\end{equation}
or
\begin{equation}\label{eq16}
\left\{\begin{aligned} &\omega_f\frac{\partial W(\rho-1)}{\partial
x}+\Omega_a(1-\rho)-\Omega_d (\rho-W)=0,\cr &c\omega_f
W\rho-(\omega_s+\omega_f)W+\omega_h (\rho-W)=0.
\end{aligned}\right.
\end{equation}
The second equation implies
\begin{equation}\label{eq17}
W=\frac{n\rho}{n+l+1-c\rho}.
\end{equation}
The steady state flux is then proportional to
\begin{equation}\label{eq18}
J=W(1-\rho)=\frac{n\rho(1-\rho)}{n+l+1-c\rho}.
\end{equation}
The same as in \cite{Nishinari2007}, at the steady state, we can
obtain the left boundary conditions
\begin{equation}\label{eq19}
S(0)=\frac{\alpha-[c\alpha(\alpha-\omega_s)/\omega_f]}{c\alpha+\omega_h},\qquad
W(0)=\frac{\alpha}{\omega_f},
\end{equation}
and the right boundary conditions
\begin{equation}\label{eq20}
S(1)=\frac{\omega_s+\beta}{\omega_h}\left[\frac{\omega_h}{\omega_h+\omega_s+\beta}-\frac{\beta}{\omega_f}\right],\qquad
W(1)=\frac{\omega_h}{\omega_h+\omega_s+\beta}-\frac{\beta}{\omega_f}.
\end{equation}

One can find that Eqs. (\ref{eq16}) involves only the
first-order derivatives of $\rho$ and $W$ with respect to $x$
whereas there are two sets of boundary conditions (\ref{eq19}) and
(\ref{eq20}). Therefore, if we integrate the equations
(\ref{eq16}) with the left boundary conditions (\ref{eq19}), the
solution (denoted by $\rho_l,\ W_l,\ S_l$ respectively) may not, in general,
match smoothly with the solution (denoted by $\rho_r,\ W_r,\
S_r$ respectively) obtained for the same equations but with the right boundary
conditions (\ref{eq20}). This discontinuity corresponds to a shock
or domain wall. However, at any position $x$, the continuity condition of motor flux, or equivalently $J_l(x)=J_r(x)$, should be satisfied, where
$J_l(x)=W(x-)[1-\rho(x-)]$ and $J_r(x)=W(x+)[1-\rho(x+)]$. At the shock position $x_s$,
\begin{equation}\label{eq21}
\left\{\begin{aligned}
& W(x_s-)=W_l(x_s),\quad \rho(x_s-)=\rho_l(x_s),\cr
& W(x_s+)=W_r(x_s),\quad \rho(x_s+)=\rho_r(x_s).
\end{aligned}\right.
\end{equation}
So the continuity condition $J_l(x_s)=J_r(x_s)$ implies
\begin{equation}\label{eq22}
\frac{n\rho_l(x_s)(1-\rho_l(x_s))}{n+l+1-c\rho_l(x_s)}=J(x_s)
=\frac{n\rho_r(x_s)(1-\rho_r(x_s))}{n+l+1-c\rho_r(x_s)},
\end{equation}
or
\begin{equation}\label{eq23}
\rho_l(x_s)+\rho_r(x_s)=1+\frac{c}{n}J(x_s)=1+\gamma\rho_l(x_s)\rho_r(x_s),
\end{equation}
where $\gamma=\frac{c}{n+l+1}<1$.

From Eqs. (\ref{eq16}), one can show the probability $\rho$ satisfies
\begin{equation}\label{eq24}
(\gamma\rho^2-2\rho+1)\rho_x
=\Omega_{ah}(1-\gamma\rho)\left[(k+1)c\rho^2-\left(\frac{c}{\gamma}+c+k(l+1)\right)\rho+\frac{c}{\gamma}\right],
\end{equation}
with $\Omega_{ah}=\Omega_{a}/\omega_{h}$. It can be proved
that, for $0\le c\le 1$ and $l\ge 0$, the discriminant
$$\begin{aligned}
\triangle :=&
\left[\frac{c}{\gamma}+c+k(l+1)\right]^2-\frac{4(k+1)c^2}{\gamma}\ge 0.
\end{aligned}
$$
So the equation (\ref{eq24}) can be reformulated as
\begin{equation}\label{eq25}
(\rho-\rho_3)(\rho-\rho_4)\rho_x=-\Omega_{ah}(k+1)c(\rho-\rho_0)(\rho-\rho_1)(\rho-\rho_2),
\end{equation}
where
\begin{equation}\label{eq26}
\begin{aligned}
&\rho_0=\frac{1}{\gamma},\qquad\rho_{3,4}=\frac{1\mp\sqrt{1-\gamma}}{\gamma},\cr
&\rho_{1,2}=\frac{\left[\frac{c}{\gamma}+c+k(l+1)\right]\mp\sqrt{\left[\frac{c}{\gamma}+c+k(l+1)\right]^2-\frac{4(k+1)c^2}{\gamma}}}{2(k+1)c}.
\end{aligned}
\end{equation}
One can easily show that, for $0\le c\le 1, l\ge 0$ and $k\ge 0$,
\begin{equation}\label{eq30}
\rho_0\ge 1,\quad\rho_4\ge 1,\quad\rho_2\ge 1,\quad \rho_1\le
1\quad\frac12\le\rho_3 \le 1.
\end{equation}
Particularly,
\begin{equation}\label{eq31}
\begin{aligned}
\rho_1=&\frac{\frac{2c}{\gamma}}{\left[\frac{c}{\gamma}+c+k(l+1)\right]+\sqrt{\left[\frac{c}{\gamma}+c+k(l+1)\right]^2-\frac{4(k+1)c^2}{\gamma}}}\cr
\to &\left\{\begin{array}{lcl}&1\qquad &\text{as\ }k\to 0,\cr
&\frac{l+n+1}{l+n+1+k(l+1)}\le 1\quad &\text{as\ }c\to
0.\end{array}\right.
\end{aligned}
\end{equation}
Moreover, one can easily show that the function
$f(x)=x-\sqrt{x^2-x}$ decreases with $x\ge 1$ monotonously. So,
for $c\le (1-k)(l+1)+n$,
\begin{equation}\label{eq32}
\begin{aligned}
\rho_1=&\frac{\left[\frac{c}{\gamma}+c+k(l+1)\right]-\sqrt{\left[\frac{c}{\gamma}+c+k(l+1)\right]^2-\frac{4(k+1)c^2}{\gamma}}}{2(k+1)c}\cr
=&\frac{(k+1)(l+1)+n+c}{2(k+1)c}-\sqrt{\left(\frac{(k+1)(l+1)+n+c}{2(k+1)c}\right)^2-\frac{l+n+1}{(k+1)c}}\cr
\ge
&\frac{(k+1)(l+1)+n+c}{2(k+1)c}-\sqrt{\left(\frac{(k+1)(l+1)+n+c}{2(k+1)c}\right)^2-\frac{(k+1)(l+1)+n+c}{2(k+1)c}}\cr
\ge
&\frac{l+n+1}{c}-\sqrt{\left(\frac{l+n+1}{c}\right)^2-\frac{l+n+1}{c}}\cr
=&\rho_3.
\end{aligned}
\end{equation}
Therefore, in the following, we always assume that $\rho_0, \rho_2, \rho_4\ge
1$, and $\frac12\le \rho_3\le \rho_1\le 1$.

The general solutions of Eq. (\ref{eq25}) are
\begin{equation}\label{eq27}
F(\rho)=x+C,
\end{equation}
where $C$ is an arbitrary constant and
\begin{equation}\label{eq28}
F(\rho)=-\frac{1}{\Omega_{ah}(k+1)c}\left[A\ln|\rho-\rho_0|+B\ln|\rho-\rho_1|+D\ln|\rho-\rho_2|\right],
\end{equation}
with
\begin{equation}\label{eq29}
A=\frac{\rho_0(\rho_0-1)}{(\rho_0-\rho_1)(\rho_2-\rho_0)},\quad
B=-\frac{\rho_1^2-2\rho_0\rho_1+\rho_0}{(\rho_0-\rho_1)(\rho_1-\rho_2)},\quad
D=\frac{\rho_2^2-2\rho_0\rho_2+\rho_0}{(\rho_0-\rho_2)(\rho_1-\rho_2)}.
\end{equation}
So the solution of Eq. (\ref{eq25}), which satisfies the left
boundary condition $\rho(0)=S(0)+W(0)$, see Eq. (\ref{eq19}), is
\begin{equation}\label{eq33}
F(\rho_l)=x+F_0,\quad \text{or}\quad \rho_l(x)=F^{-1}(x+F_0),
\end{equation}
where $F_0=F[\rho(0)]$. Similarly, the solution of Eq.
(\ref{eq25}), which satisfies the right boundary condition
$\rho(1)=S(1)+W(1)$, see Eq. (\ref{eq20}), is
\begin{equation}\label{eq34}
F(\rho_r)=x+F_1-1,\quad \text{or}\quad \rho_r(x)=F^{-1}(x+F_1-1),
\end{equation}
where $F_1=F(\rho(1))$. In the following, we assume
$\rho_l\not\equiv\rho_r$ (otherwise, there would be no shock and
boundary layers).

Combining (\ref{eq21}) (\ref{eq23}) (\ref{eq33}) (\ref{eq34}), one sees that, at
the shock position $x_s$
\begin{equation}\label{eq35}
\left\{\begin{aligned}
&F(\rho_l(x_s))-F\left(\rho_r(x_s)\right)+F_1-F_0-1=0,\cr
&\rho_l(x_s)+\rho_r(x_s)=1+\gamma\rho_l(x_s)\rho_r(x_s).
\end{aligned}\right.
\end{equation}
These are the efficient and necessary condition of the existence of
shock at the position $x_s$. In other words, at the shock position
$x_s$, $H(x_s,\Omega_a,k,\omega_f,l,n,c):=$
\begin{equation}\label{eq36}
\left.\begin{aligned}
F(\rho_l(x_s))-F\left(\frac{1-\rho_l(x_s)}{1-\gamma\rho_l(x_s)}\right)+F_1-F_0-1=0.
\end{aligned}\right.
\end{equation}
If there exists $0<x_s<1$, such that
$H(x_s,\Omega_a,k,\omega_f,l,n,c)=0$, the shock will appear at
$x_s$, and the height of the shock is
\begin{equation}\label{eq37}
\left.\begin{aligned}
\varepsilon_s=|\rho_r(x_s)-\rho_l(x_s)|=\left|\frac{1-2\rho_l(x_s)+\gamma
\rho_l^2(x_s)}{1-\gamma\rho_l(x_s)}\right|.
\end{aligned}\right.
\end{equation}

\begin{figure}
  \includegraphics[width=200pt]{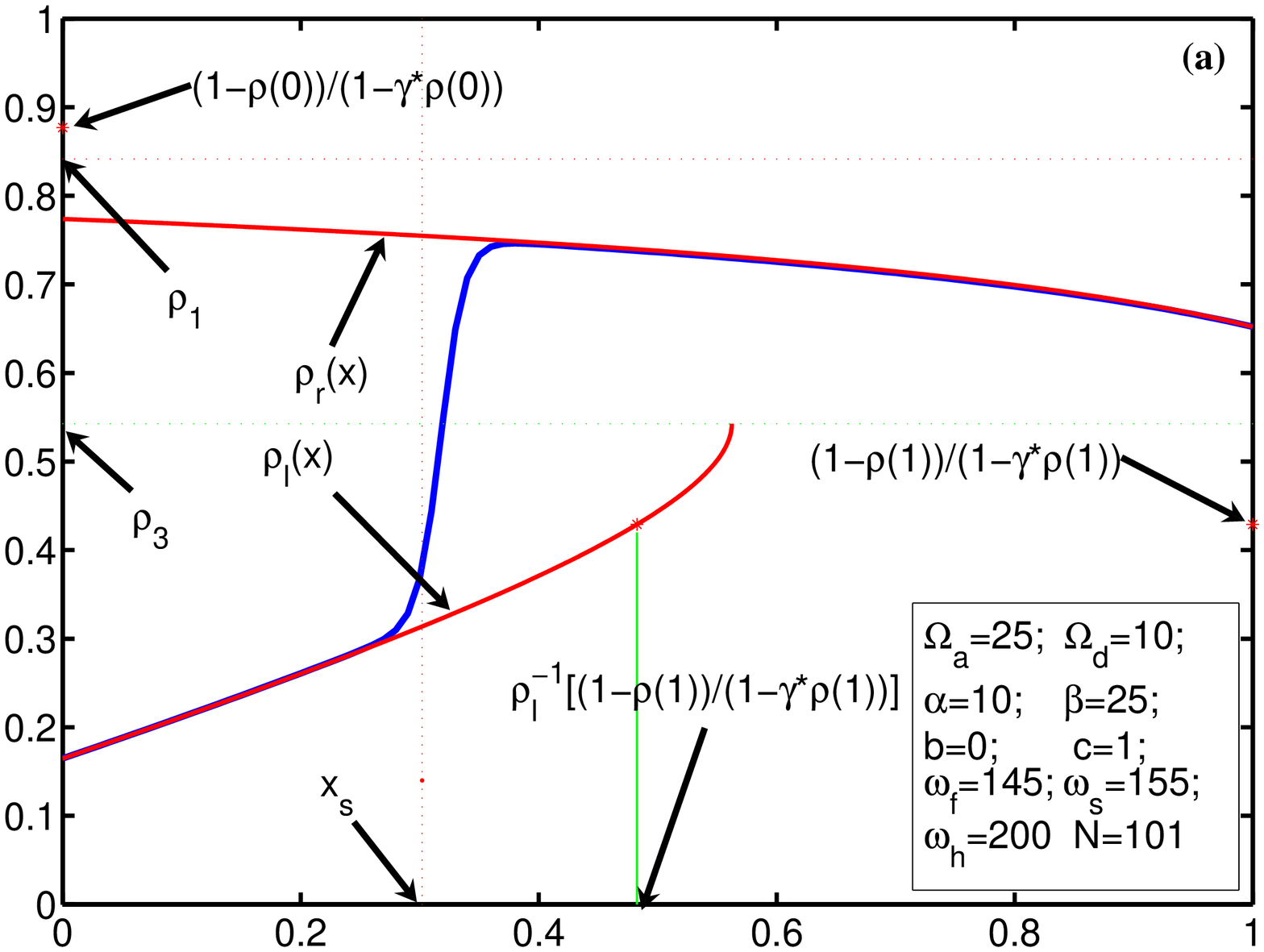}\includegraphics[width=200pt]{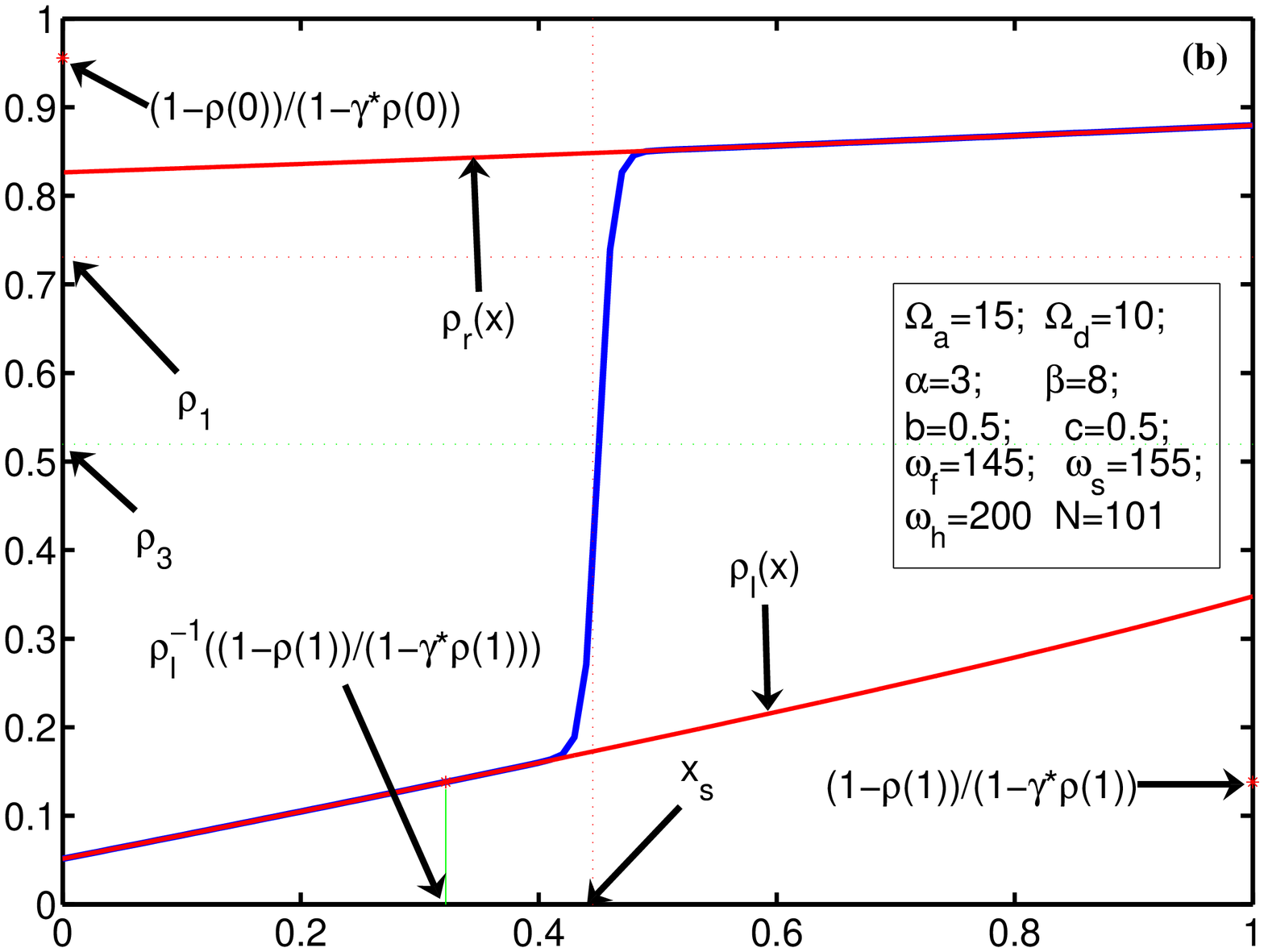}
  \includegraphics[width=200pt]{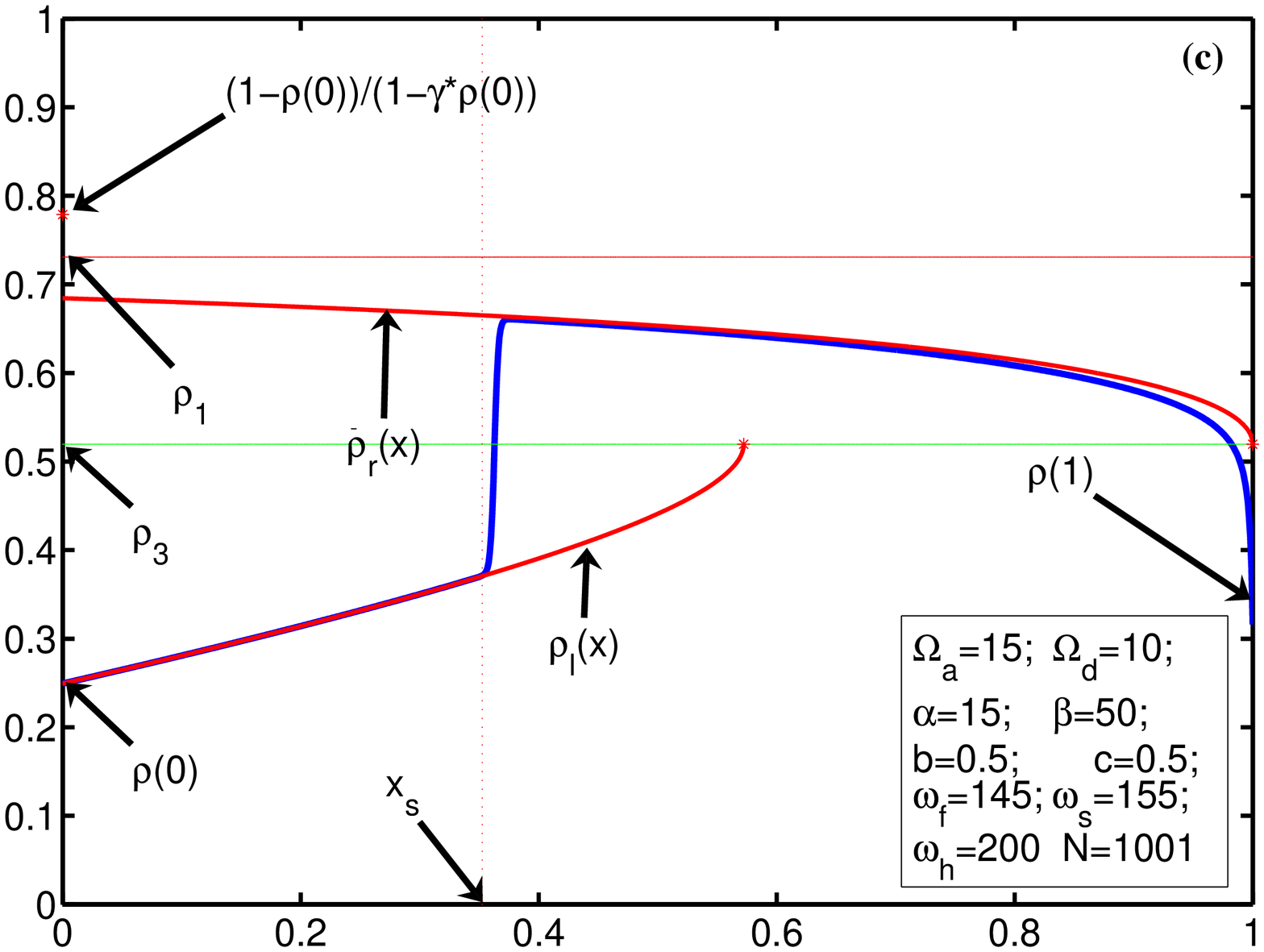}\includegraphics[width=200pt]{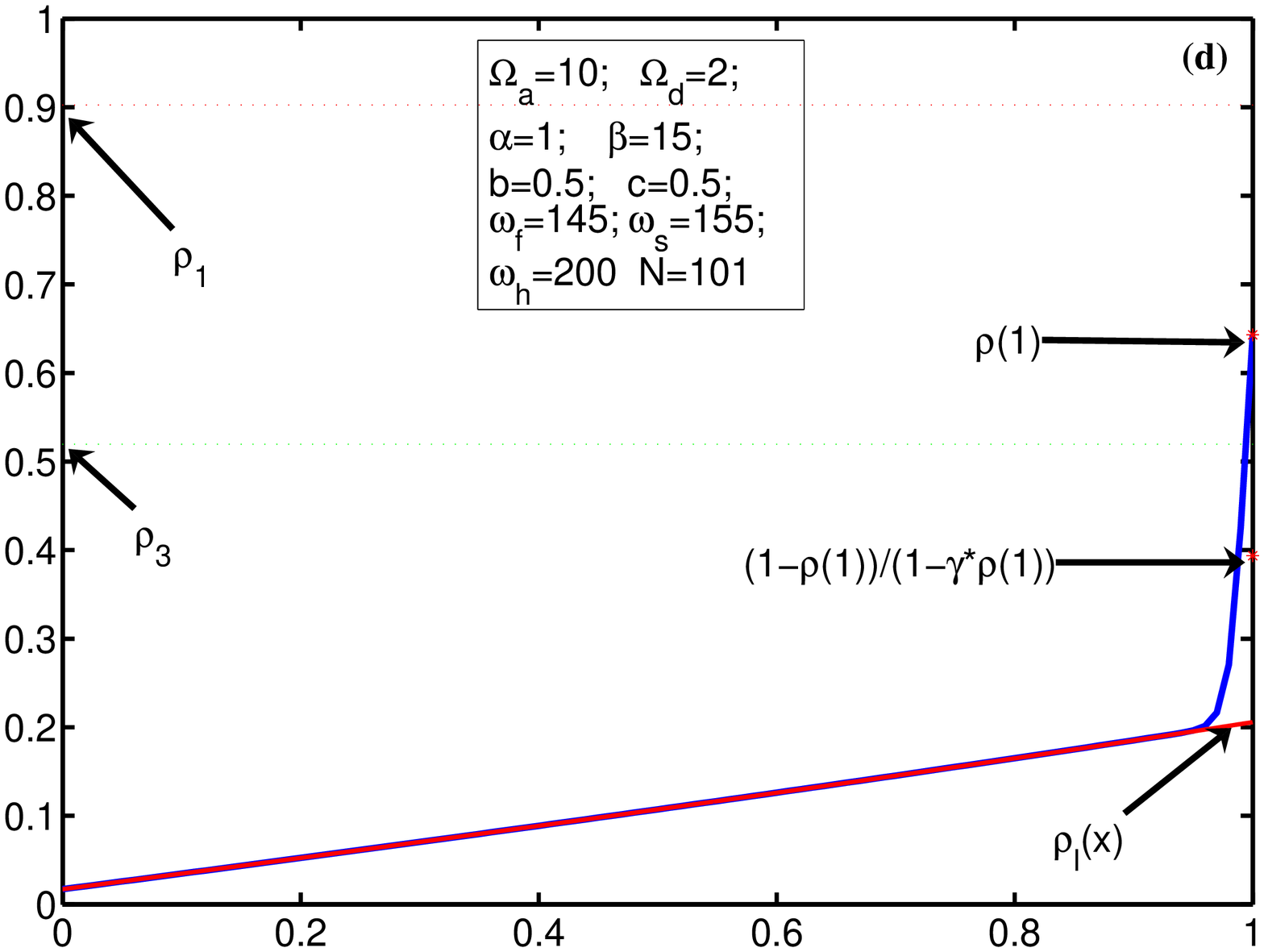}
  \caption{There exists shock in $(0, 1)$ if $0\le\rho(0)<\rho_3$, $\rho_3<\rho(1)\le
\rho_1$ and $F\left(\frac{1-\rho(1)}{1-\gamma\rho(1)}\right)<F_0+1$,
$F\left(\frac{1-\rho(0)}{1-\gamma\rho(0)}\right)<F_1-1$ {\bf (a)}; Or
  $0\le\rho(0)<\rho_3$,
$\rho_1\le\rho(1)\le 1$ and
$F\left(\frac{1-\rho(1)}{1-\gamma\rho(1)}\right)<F_0+1$,
$F\left(\frac{1-\rho(0)}{1-\gamma\rho(0)}\right)>F_1-1$  {\bf (b)}.
  There exists shock in $(0,1)$ and boundary layer at $x=1$ if $0\le \rho(0), \rho(1)\le \rho_3$ and $\rho_l^{-1}(\rho_3)<1$,
$\tilde{\rho}_r(0)<\frac{1-\rho(0)}{1-\gamma\rho(0)}$ {\bf (c)}. There exists boundary layer at $x=1$ if $0\le \rho(0)\le \rho_3, \rho_3\le \rho(0)\le \rho_1$ and $F\left(\frac{1-\rho(1)}{1-\gamma\rho(1)}\right)>F_0+1$ {\bf (d)}.
  }\label{Fig12}
\end{figure}

Let
\begin{equation}\label{eq361}
\left.\begin{aligned}
H(\rho):=F(\rho)-F\left(\frac{1-\rho}{1-\gamma\rho}\right)+F_1-F_0-1=0,
\end{aligned}\right.
\end{equation}
then from (\ref{eq28}), one can obtain
\begin{equation}\label{eq362}
\left.\begin{aligned} \frac{d H(\rho)}{d\rho}&=\frac{d
F(\rho)}{d\rho}-\frac{d}{d \rho}
F\left(\frac{1-\rho}{1-\gamma\rho}\right)
=\frac{\gamma\rho^2-2\rho+1}{\Omega_{ah}(1-\gamma\rho)}\cr &\times
\frac{[(1-\rho)-\rho_1(1-\gamma\rho)][(1-\rho)-\rho_2(1-\gamma\rho)]+(\gamma-1)(\rho-\rho_1)(\rho-\rho_2)}{(\rho-\rho_1)(\rho-\rho_2)[(1-\rho)-\rho_1(1-\gamma\rho)][(1-\rho)-\rho_2(1-\gamma\rho)]}\cr
&=\frac{\gamma
(1-\rho)^2(1-\gamma\rho_1)(\rho-\rho_3)(\rho-\rho_4)\left(\frac{1-\rho_1}{1-\gamma\rho_1}-\rho_2\right)}{\Omega_{ah}(1-\gamma\rho)^3(\rho-\rho_1)(\rho-\rho_2)\left(\frac{1-\rho}{1-\gamma\rho}-\rho_1\right)\left(\frac{1-\rho}{1-\gamma\rho}-\rho_2\right)}\cr
&\left\{\begin{array}{lcl}>0\qquad &\text{if}\quad &
0\le\rho<\frac{1-\rho_1}{1-\gamma\rho_1},\cr <0\qquad
&\text{if}\quad & \frac{1-\rho_1}{1-\gamma\rho_1}<\rho <\rho_1,\cr
>0\qquad &\text{if}\quad & \rho_1<\rho\le 1.
\end{array}\right.\end{aligned}\right.
\end{equation}
Using this property of the function $H(\rho)$, we can obtain the
following results:

\noindent{\bf (I)} For $0\le\rho(0)<\rho_3$ and $\rho_3<\rho(1)\le
\rho_1$, the conditions of existence of shock in interval $(0,\
1)$ is
\begin{equation}\label{eq38}
F\left(\frac{1-\rho(1)}{1-\gamma\rho(1)}\right)<F_0+1\quad
\text{and}\quad
F\left(\frac{1-\rho(0)}{1-\gamma\rho(0)}\right)<F_1-1,
\end{equation}
see Fig. \ref{Fig12} {\bf (a)}.
From (\ref{eq25}) (\ref{eq28}), one can find the function $F(\rho)$
increases with $\rho$ for $0\le \rho< \rho_3$ and $\rho_1<
\rho\le 1$, and decreases with $\rho$ for $\rho_3< \rho< \rho_1$.
Thus
\begin{equation}\label{eq39}
\begin{aligned}
&F\left(\frac{1-\rho(1)}{1-\gamma\rho(1)}\right)<F_0+1\Longleftrightarrow\rho_l^{-1}\left(\frac{1-\rho(1)}{1-\gamma\rho(1)}\right)<1,\cr
&F\left(\frac{1-\rho(0)}{1-\gamma\rho(0)}\right)<F_1-1
\Longleftrightarrow\rho_r(0)<\frac{1-\rho(0)}{1-\gamma\rho(0)}.
\end{aligned}
\end{equation}
Therefore, the conditions presented in (\ref{eq38}) are generalizations of the ones obtained in \cite{Zhang2007} for the usual TASEP with motor detachment and attachment:
\begin{equation}\label{eq43}
\begin{aligned}
\rho_l^{-1}(1-\rho(1))<1,\quad\text{and}\quad \rho_r(0)<1-\rho(0).
\end{aligned}
\end{equation}
In fact, if the parameter $c=0$ (i.e. $\gamma=0$), the Eq.
(\ref{eq24}) reduces to
\begin{equation}\label{eq44}
\begin{aligned}
(1-2\rho)\rho_x=\Omega_{ah}[(l+n+1)-[(l+n+1)+k(l+1)]\rho],
\end{aligned}
\end{equation}
which is similar as the model discussed in \cite{Parmeggiani2003} for the usual TASEP. For such reduced cases, $\rho_3=0.5$, and the conditions (\ref{eq38}) of existence of shock is reduced to (\ref{eq43}).

\noindent{\bf (II)} For $0\le\rho(0)<\rho_3$ and
$\rho_1\le\rho(1)\le 1$, the conditions of existence of shock
in interval $(0,\ 1)$ is
\begin{equation}\label{eq47}
F\left(\frac{1-\rho(1)}{1-\gamma\rho(1)}\right)<F_0+1,\quad
\text{and}\quad
F\left(\frac{1-\rho(0)}{1-\gamma\rho(0)}\right)>F_1-1,
\end{equation}
see Fig. \ref{Fig12} {\bf (b)}. Similar as in {\bf (I)},
\begin{equation}\label{eq48}
\begin{aligned}
&F\left(\frac{1-\rho(1)}{1-\gamma\rho(1)}\right)<F_0+1\Longleftrightarrow\rho_l^{-1}\left(\frac{1-\rho(1)}{1-\gamma\rho(1)}\right)<1,\cr
&F\left(\frac{1-\rho(0)}{1-\gamma\rho(0)}\right)>F_1-1\Longleftrightarrow\rho_r^{-1}\left(\frac{1-\rho(0)}{1-\gamma\rho(0)}\right)>0.
\end{aligned}
\end{equation}
Therefore, conditions (\ref{eq47}) are also generalizations of the ones for the usual TASEP \cite{Zhang2007}:
\begin{equation}\label{eq52}
\begin{aligned}
\rho_l^{-1}(1-\rho(1))<1,\quad\text{and}\quad
\rho_r^{-1}(1-\rho(0))>0.
\end{aligned}
\end{equation}

\noindent{\bf (III)} For $0\le \rho(0), \rho(1)\le \rho_3$, the
condition of the existence of shock in $(0, 1)$ is
\begin{equation}\label{eq55}
\begin{aligned}
\rho_l^{-1}(\rho_3)<1\qquad\text{and}\qquad
\tilde{\rho}_r(0)<\frac{1-\rho(0)}{1-\gamma\rho(0)}
\end{aligned}
\end{equation}
where $\tilde{\rho}_r(x)$ is one of the solutions of differential
equation (\ref{eq24}), which satisfies $\tilde{\rho}_r(1)=\rho_3$
and $\tilde{\rho}_r(0)>\rho_3$. See Fig. \ref{Fig12} {\bf (c)}.

\noindent{\bf (IV)} For $\rho(0), \rho(1)>\rho_3$, there exists no
shock in $(0, 1)$. It can be readily verified that the function
$f(\rho_l, \rho_r)=\rho_l+\rho_r-\gamma\rho_l\rho_r-1$ increases
monotonously with $0\le \rho_l, \rho_r\le 1$, and
$f(\rho_3, \rho_3)=0$. For $\rho(0), \rho(1)>\rho_3$, one knows
that $\rho_3\le\min (\rho_l, \rho_r)<\max (\rho_l, \rho_r)$ (note:
we always assume $\rho_l\not\equiv\rho_r$). Thus $f(\rho_l,
\rho_r)=\rho_l+\rho_r-\gamma\rho_l\rho_r-1>0$. It is to say that
there exists no shock [see (\ref{eq35})].

\noindent{\bf (V)} For $0\le \rho(1)< \rho_3$ and
$\rho_l(0)>\rho_3$, there exists no shock in $(0, 1)$.

In conclusion, $\rho(0)<\rho_3$ is one necessary condition of the existence of shock in $(0, 1)$.

\section{The existence of boundary layer}
\begin{figure}
  \includegraphics[width=200pt]{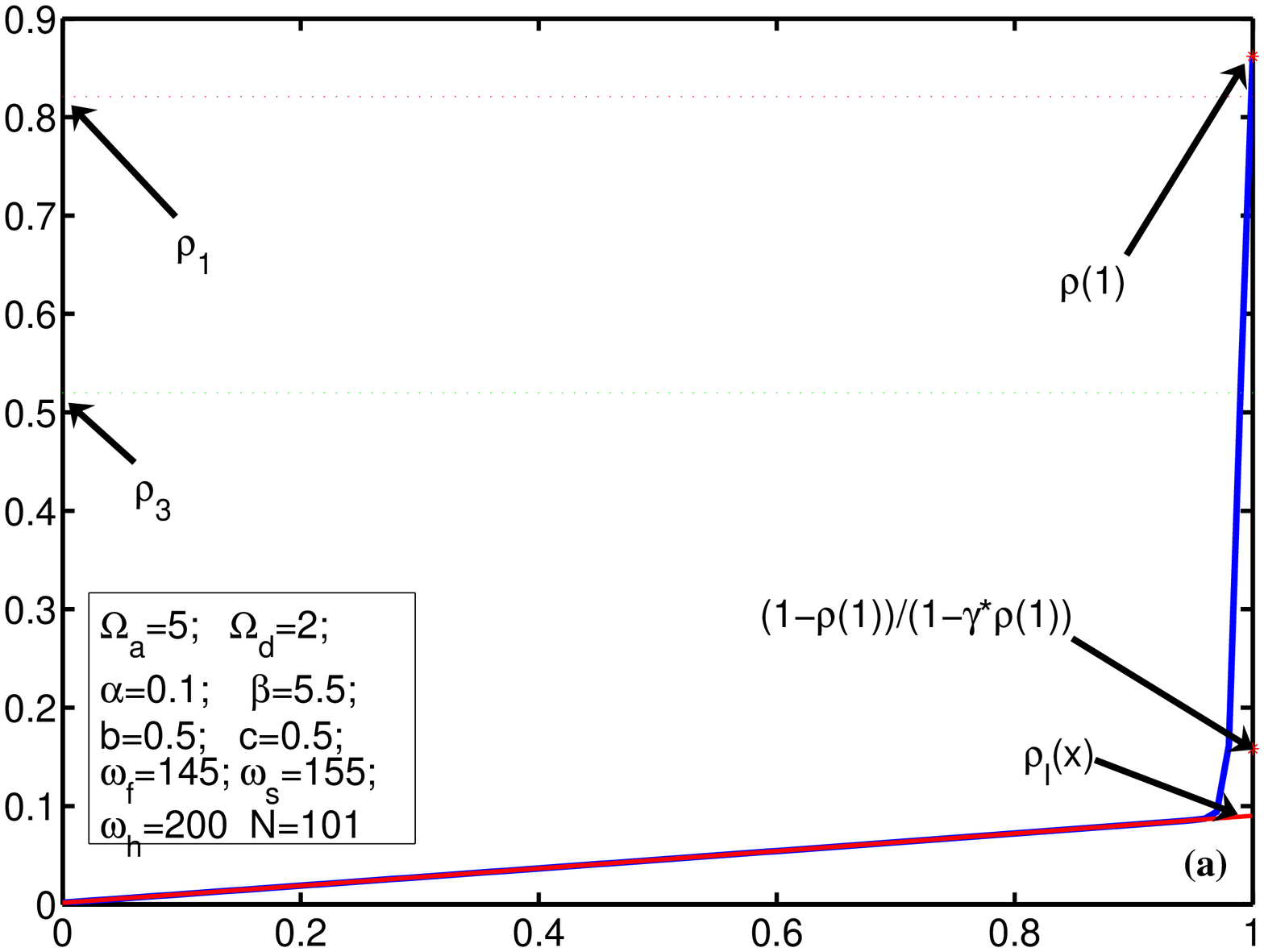}\includegraphics[width=200pt]{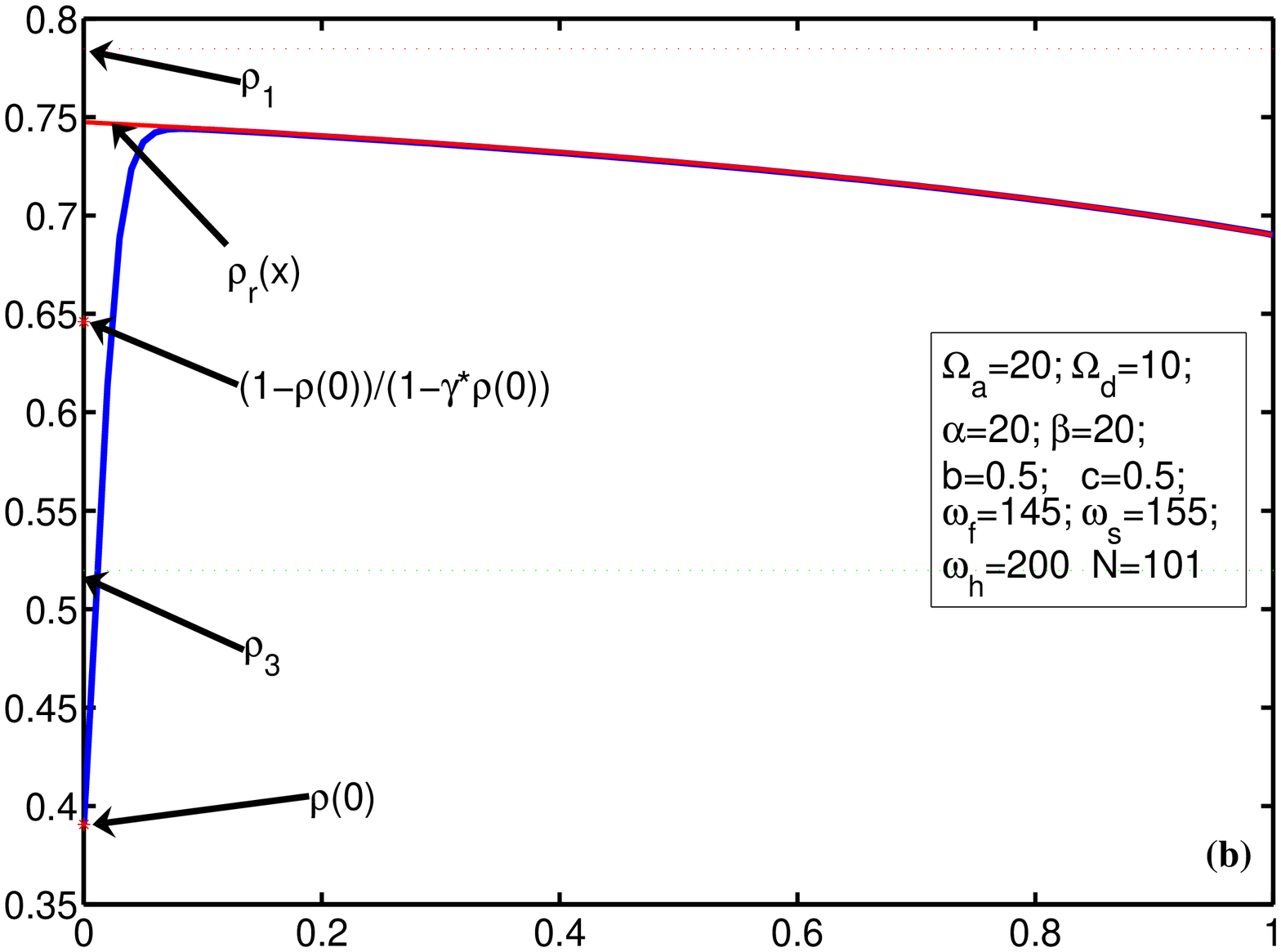}
  \includegraphics[width=200pt]{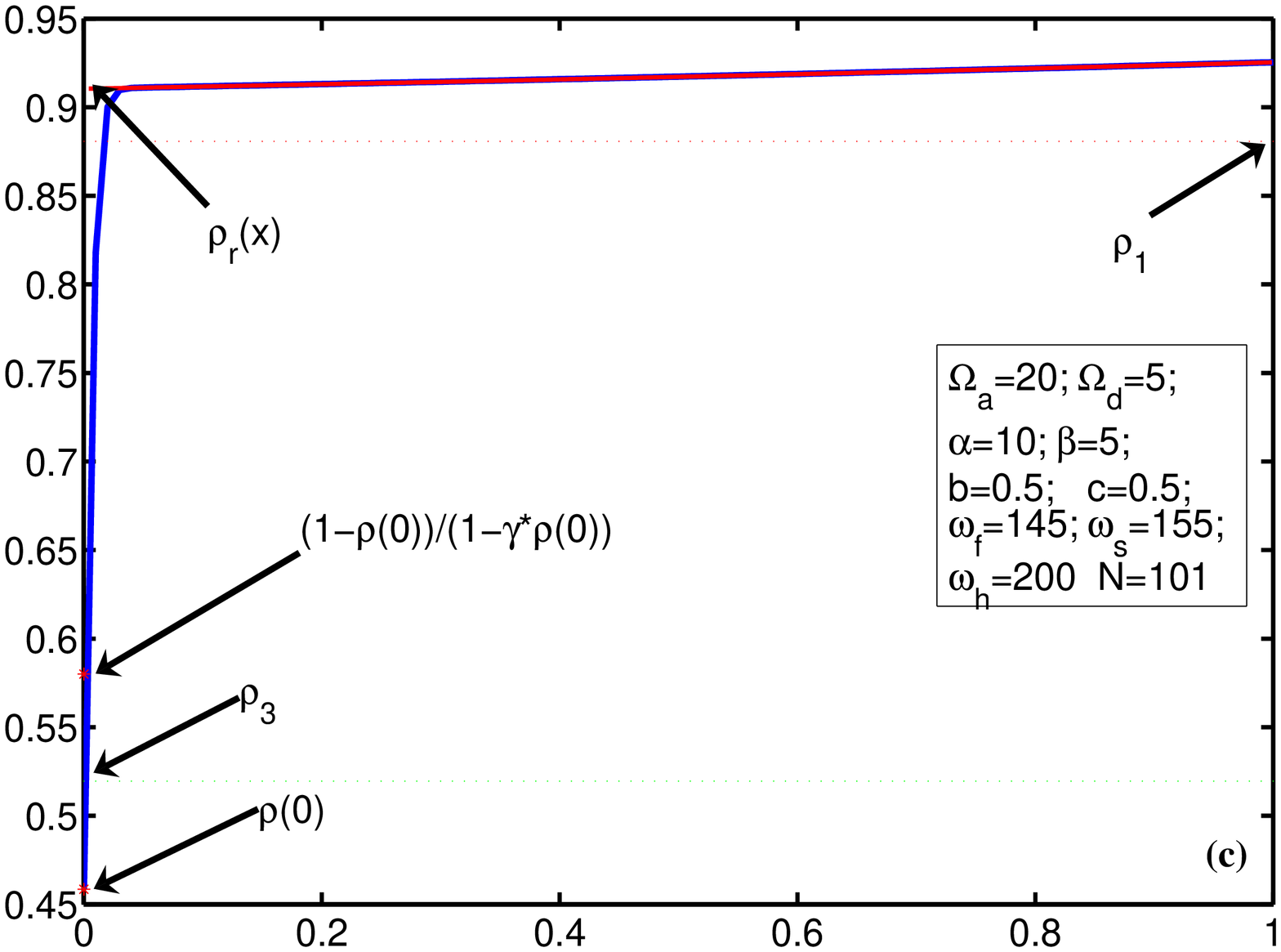}\includegraphics[width=200pt]{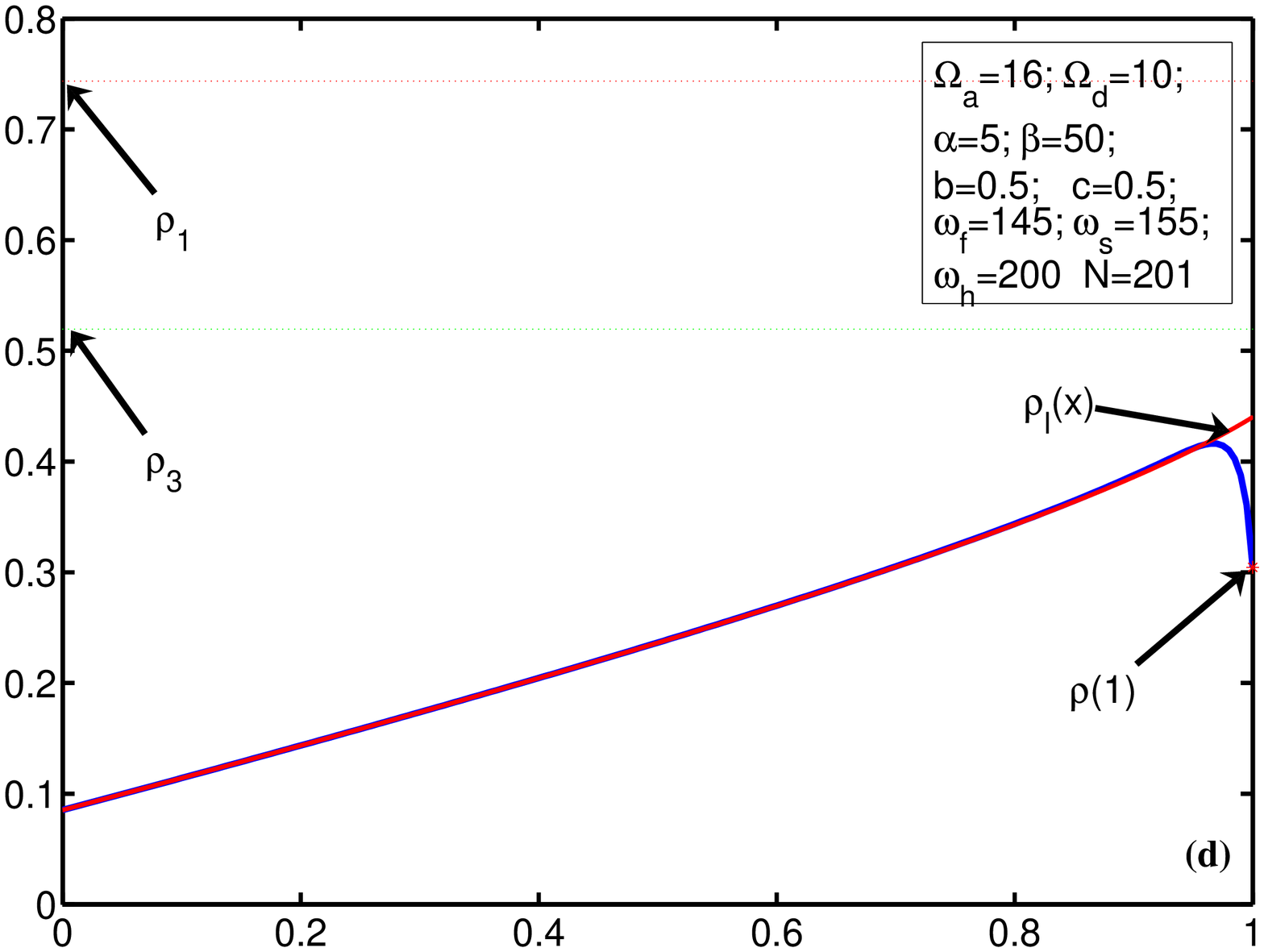}
  \caption{There exists boundary layers at $x=1$ if $0\le \rho(0)\le \rho_3, \rho_1\le \rho(1)\le 1$ and
  $F\left(\frac{1-\rho(1)}{1-\gamma\rho(1)}\right)>F_0+1$ {\bf (a)}; Or $0\le \rho(0), \rho(1)\le \rho_3$ and
$\rho_l(1)> \rho(1)$ {\bf (d)}.
  There exists boundary layers at $x=0$ if $0\le \rho(0)<\rho_3, \rho_3<\rho(1)\le \rho_1$ and $F\left(\frac{1-\rho(0)}{1-\gamma\rho(0)}\right)>F_1-1$ {\bf (b)}; Or $0\le \rho(0)\le \rho_3, \rho_1\le \rho(1)\le 1$ and
  $F\left(\frac{1-\rho(0)}{1-\gamma\rho(0)}\right)<F_1-1$ {\bf (c)}.}\label{Fig34}
\end{figure}

Generally speaking, if $\rho_l\not\equiv\rho_r$ and there is no shock
in $(0, 1)$, boundary layer will appear at least at one of
the boundaries 0 and 1. Similar as in \cite{Zhang2007}, we have the
following results:

\noindent{\bf (I)} For $0\le \rho(0)<\rho_3$ and $\rho_3<\rho(1)\le
\rho_1$: if
\begin{equation}\label{eq60}
F\left(\frac{1-\rho(1)}{1-\gamma\rho(1)}\right)>F_0+1\quad
\text{and}\quad
F\left(\frac{1-\rho(0)}{1-\gamma\rho(0)}\right)<F_1-1,
\end{equation}
there exists boundary layer at the right boundary $x=1$, see Fig.
\ref{Fig12} {\bf (d)}; if
\begin{equation}\label{eq61}
F\left(\frac{1-\rho(1)}{1-\gamma\rho(1)}\right)<F_0+1\quad
\text{and}\quad
F\left(\frac{1-\rho(0)}{1-\gamma\rho(0)}\right)>F_1-1,
\end{equation}
there exists boundary layer at the left boundary $x=0$ (in these
cases, the shock position $x_s<0$), see Fig. \ref{Fig34} {\bf
(b)}.

In view of the property (\ref{eq362}) of function $H(\rho)$,
the conditions (\ref{eq60}) can be simplified as
\begin{equation}\label{eq601}
F\left(\frac{1-\rho(1)}{1-\gamma\rho(1)}\right)>F_0+1,
\end{equation}
and the conditions (\ref{eq61}) can be simplified as
\begin{equation}\label{eq611}
F\left(\frac{1-\rho(0)}{1-\gamma\rho(0)}\right)>F_1-1.
\end{equation}

\noindent{\bf (II)} For $0\le\rho(0)<\rho_3$ and
$\rho_1\le\rho(1)\le 1$: if
\begin{equation}\label{eq62}
F\left(\frac{1-\rho(0)}{1-\gamma\rho(0)}\right)<F_1-1,
\end{equation}
there exists boundary layer at the left boundary $x=0$ (i.e. the
shock position $x_s<0$), see Fig. \ref{Fig34} {\bf (c)}; if
\begin{equation}\label{eq63}
F\left(\frac{1-\rho(1)}{1-\gamma\rho(1)}\right)>F_0+1,
\end{equation}
there exists boundary layer at the right boundary $x=1$ (i.e. the
shock position $x_s>1$), see Fig. \ref{Fig34} {\bf (a)}.

\begin{figure}
  \includegraphics[width=200pt]{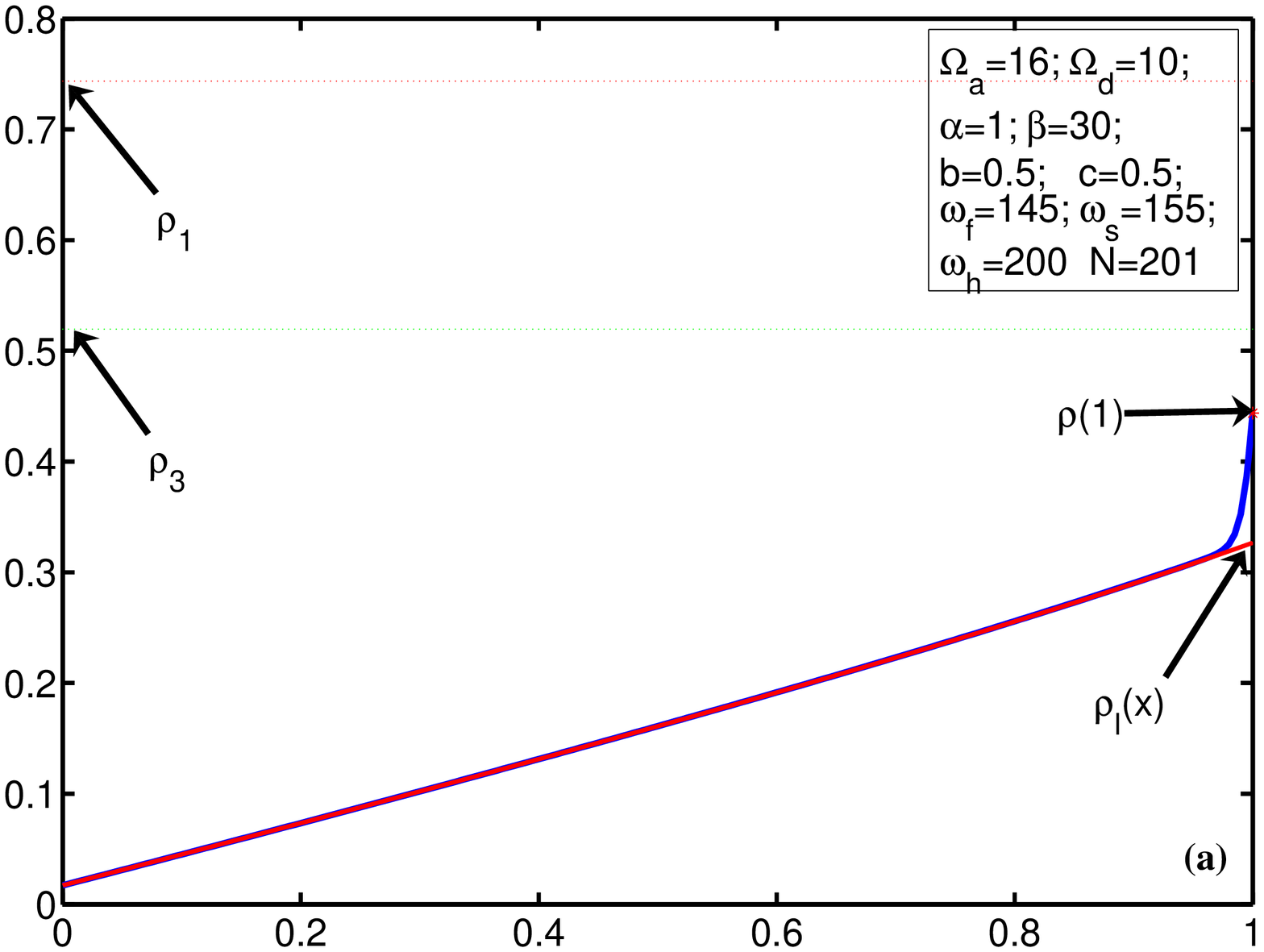}\includegraphics[width=200pt]{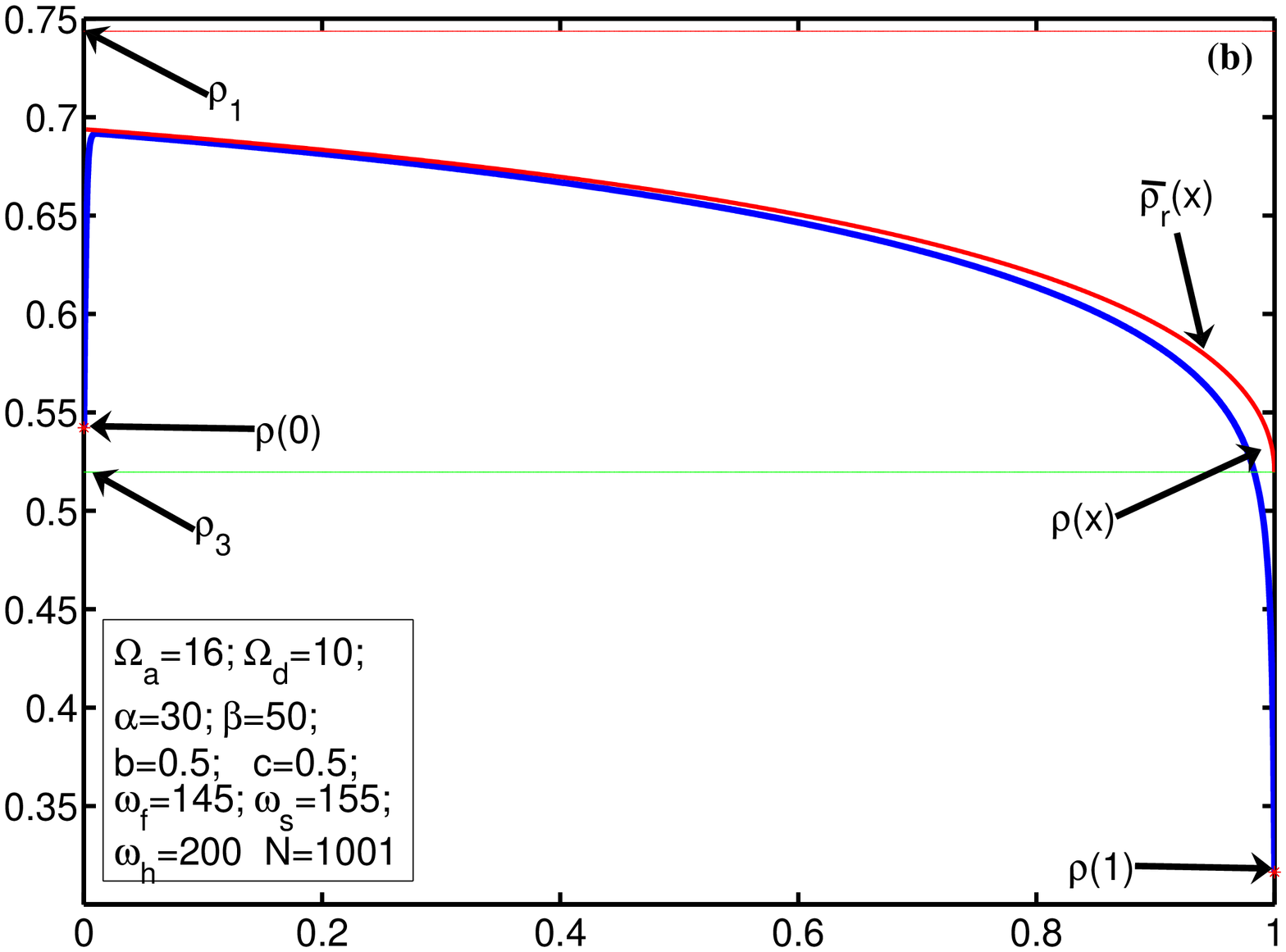}
  \includegraphics[width=200pt]{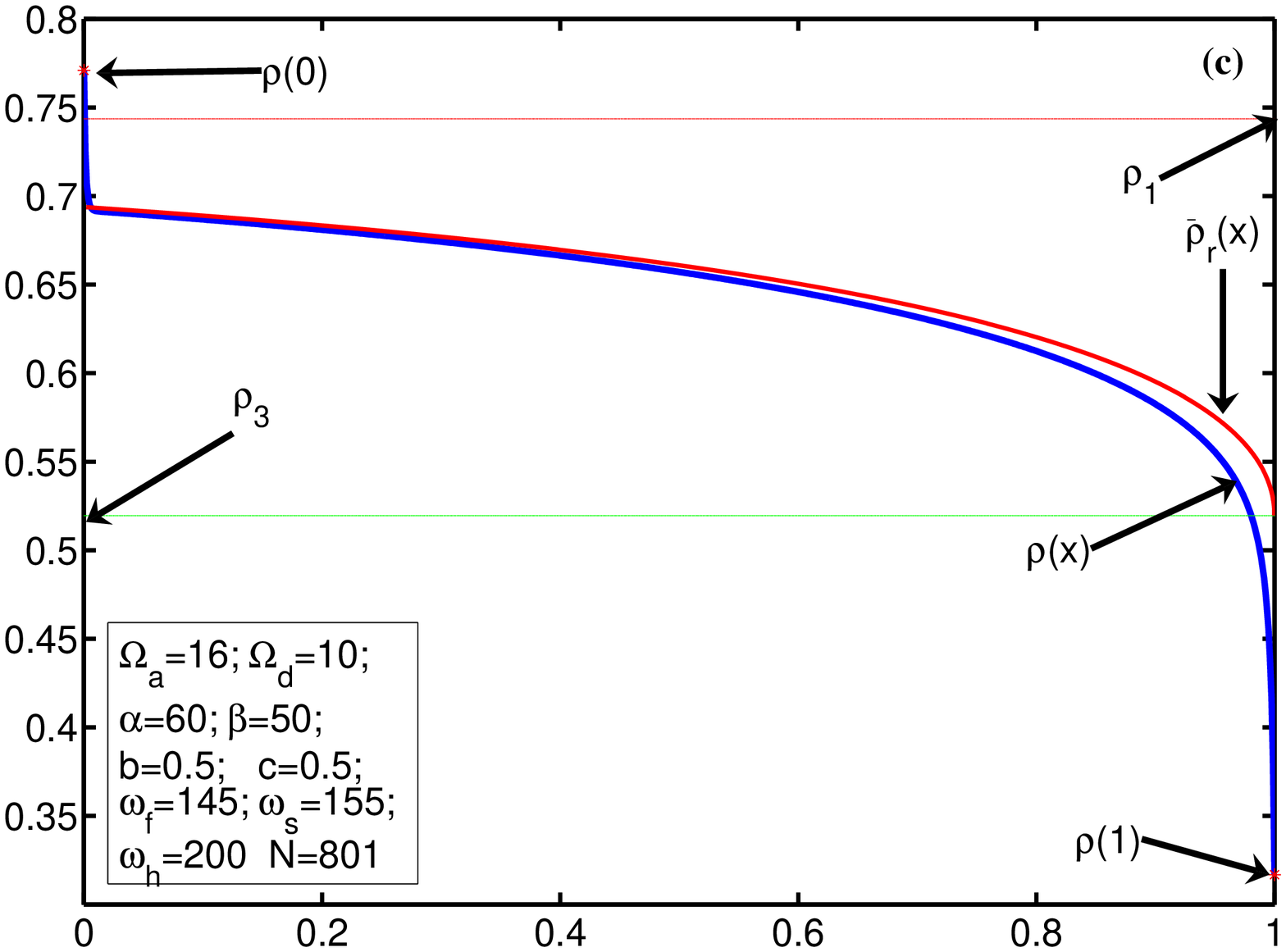}\includegraphics[width=200pt]{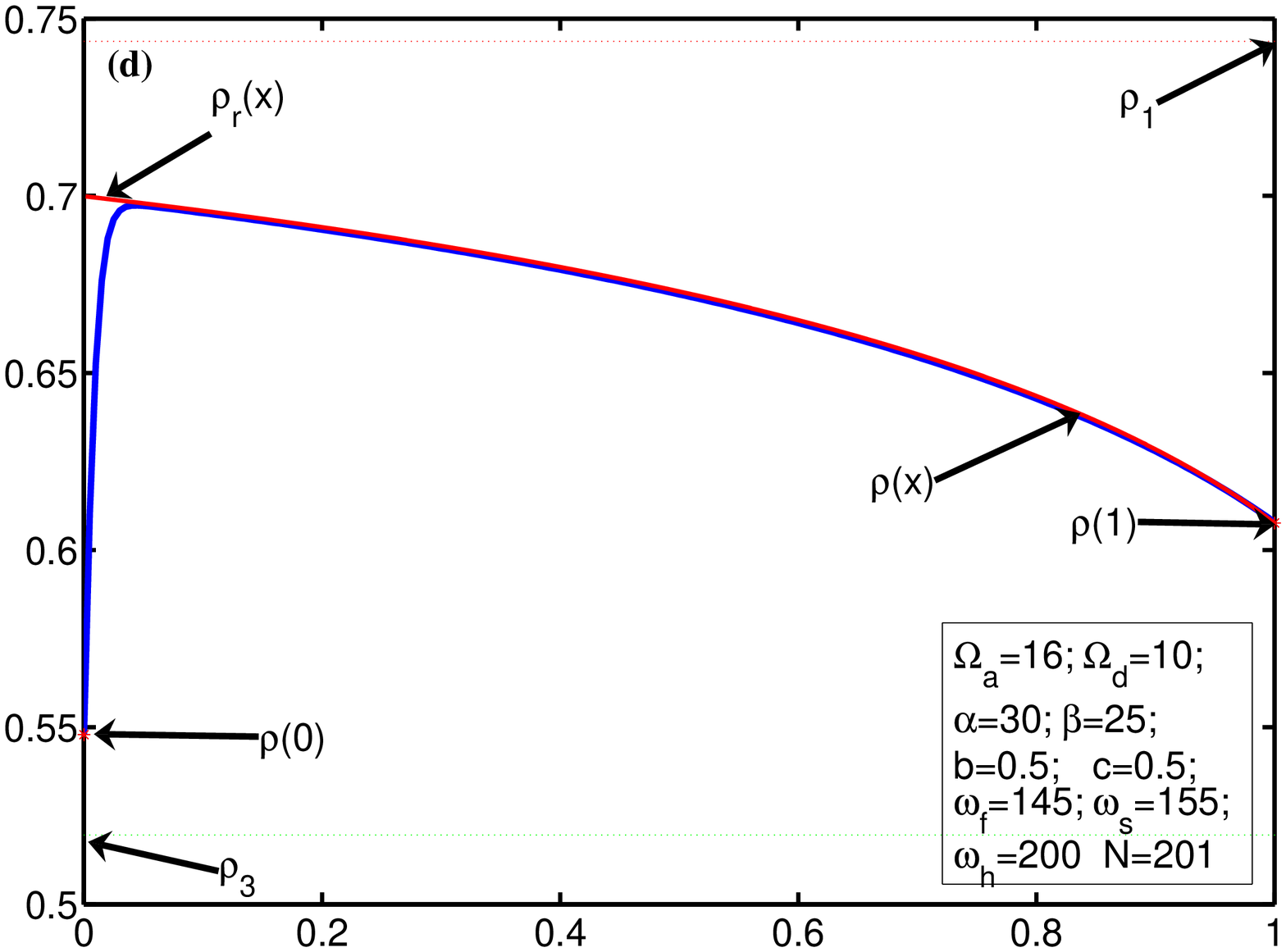}
  \caption{There exists boundary layers at $x=1$ if $0\le \rho(0), \rho(1)\le \rho_3$ and
$\rho_l(1)< \rho(1)$ {\bf (a)}. There exist boundary layers at both $x=0$ and $x=1$ if $\rho_3\le \rho(0)<1, 0\le\rho(1)\le \rho_3$
and $\rho(0)<\tilde{\rho}_r(0)$ {\bf (b)}; Or
 $\rho_3\le \rho(0)<\rho_1, 0\le\rho(1)\le \rho_3$ and $\rho(0)>\tilde{\rho}_r(0)$ {\bf (c)}. There are boundary layers at $x=0$ if $\rho_3\le \rho(0), \rho(1)\le
\rho_1$ and $\rho(0)< \rho_r(0)$ {\bf (d)}.
}\label{Fig56}
\end{figure}

\noindent{\bf (III)} For $0\le \rho(0), \rho(1)\le \rho_3$ and
$\rho_l(1)\ne \rho(1)$, there exists boundary layer at $x=1$. See
 Fig. \ref{Fig12} {\bf (c)}, Fig. \ref{Fig34} {\bf (d)} and  Fig. \ref{Fig56} {\bf (a)}. If
\begin{equation}\label{eq64}
\begin{aligned}
\rho_l^{-1}(\rho_3)<1\quad\text{and}\quad
\tilde{\rho}_r(0)>\frac{1-\rho(0)}{1-\gamma\rho(0)},
\end{aligned}
\end{equation}
there is also the boundary layer at the left boundary $x=0$.

\noindent{\bf (IV)} For $\rho_3\le \rho(0)<1, 0\le\rho(1)\le \rho_3$
and $\rho(0)\ne\tilde{\rho}_r(0)$, there exist boundary layers at
both $x=0$ and $x=1$, see
Fig. \ref{Fig56} {\bf (b)} and Fig. \ref{Fig56} {\bf (c)}. For these cases, $\rho(x)=\tilde{\rho}_r(x)$ for $0<x<1$.

\begin{figure}
  \includegraphics[width=200pt]{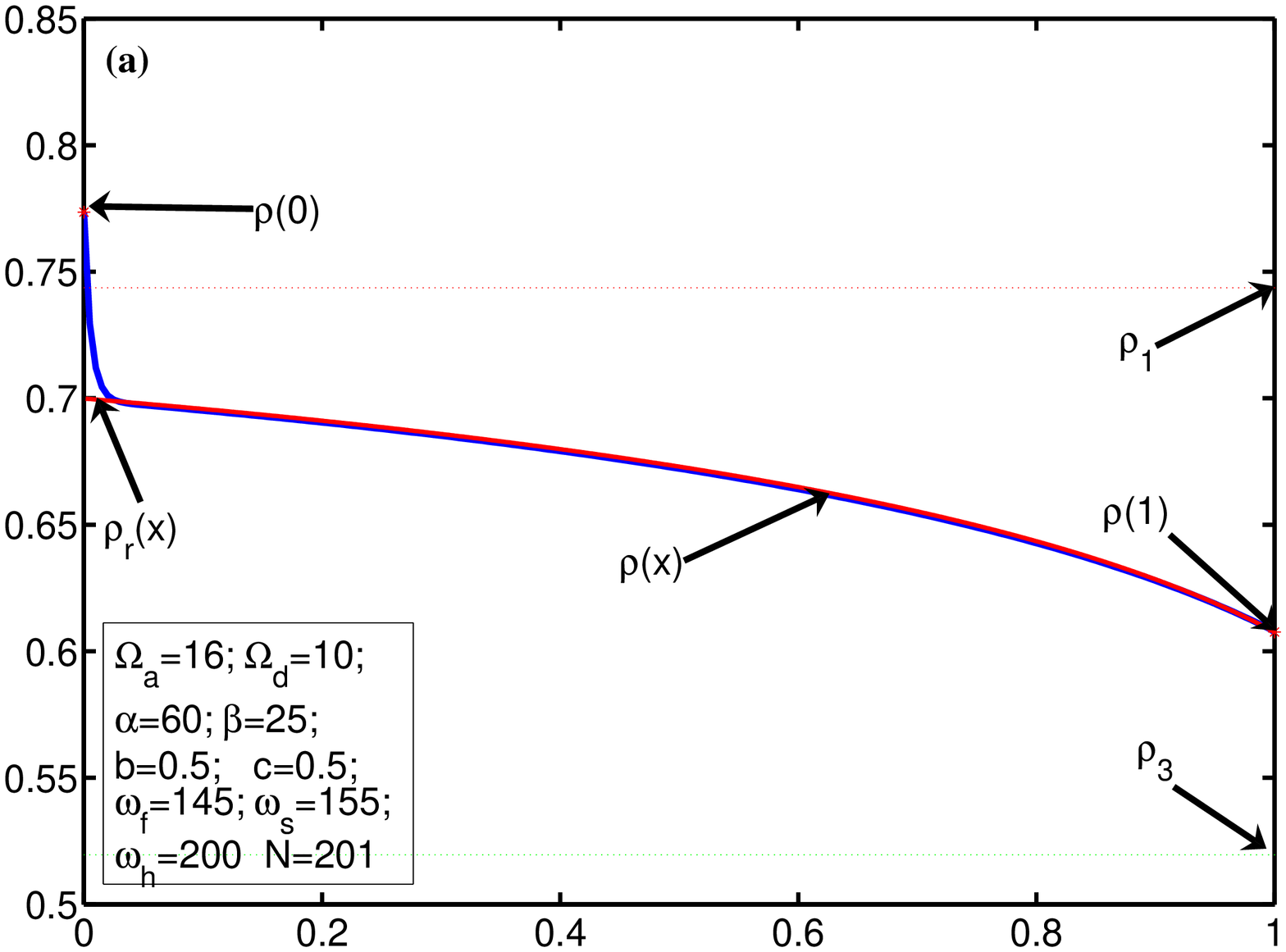}\includegraphics[width=200pt]{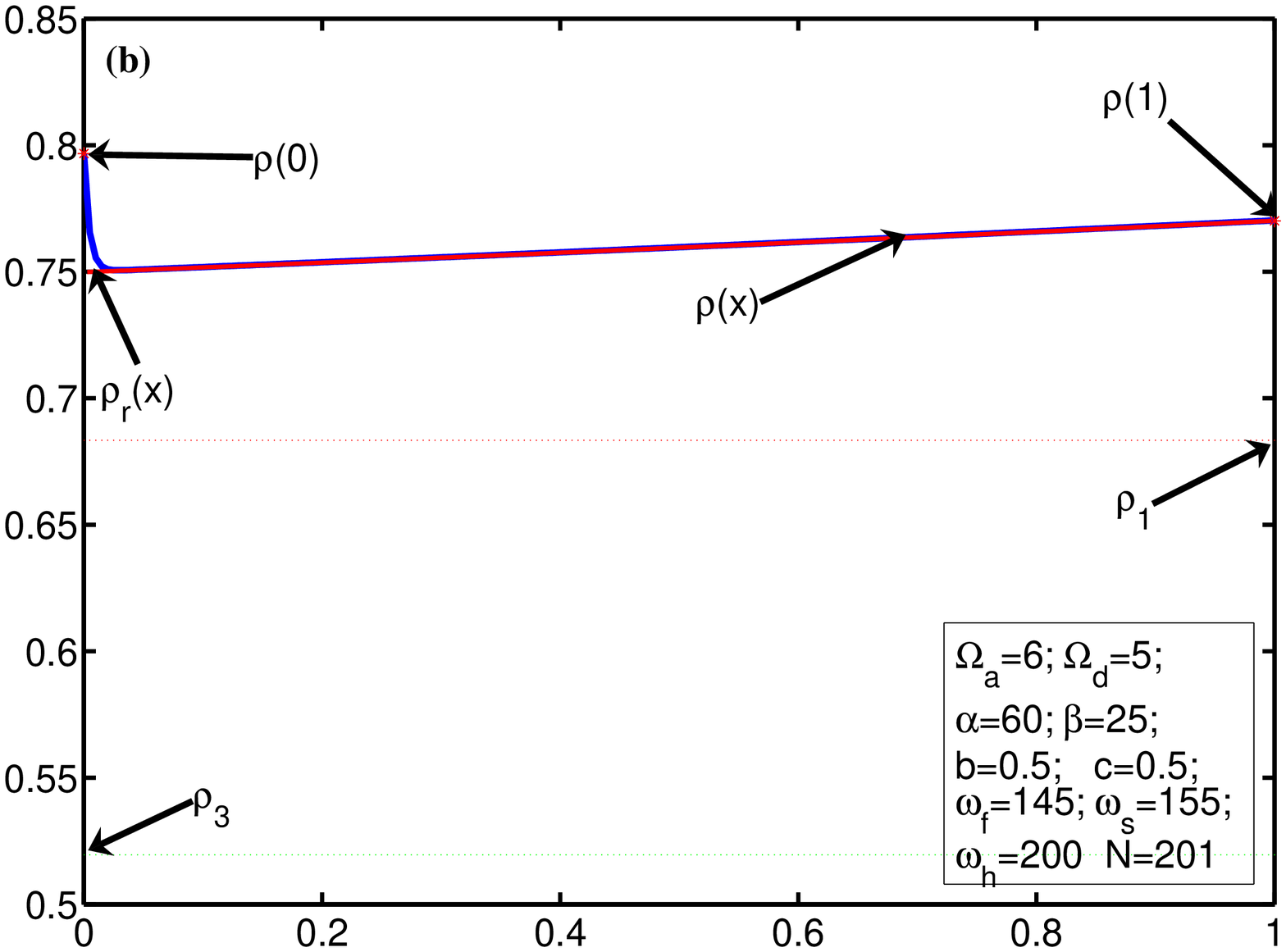}
  \includegraphics[width=200pt]{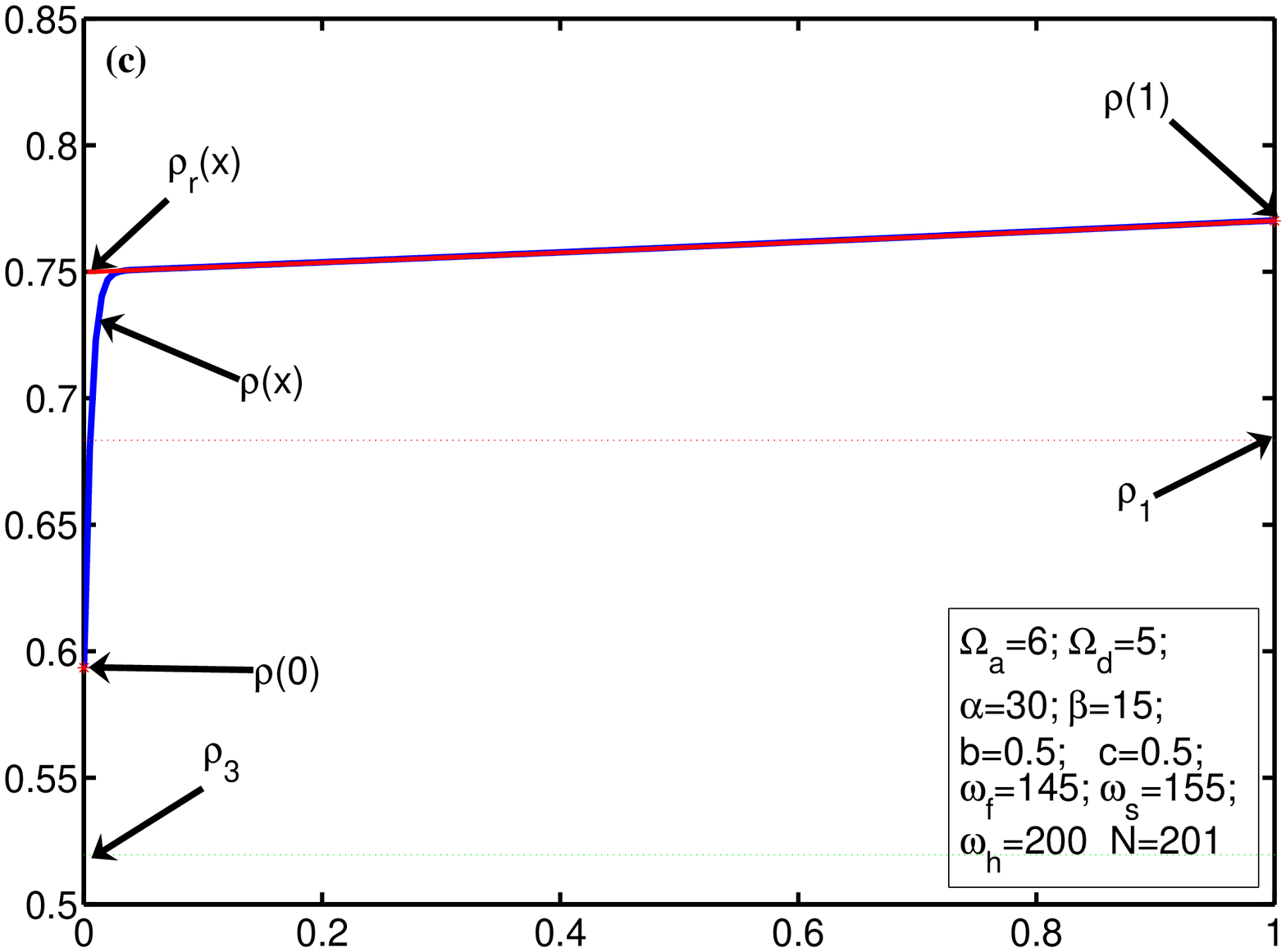}\includegraphics[width=200pt]{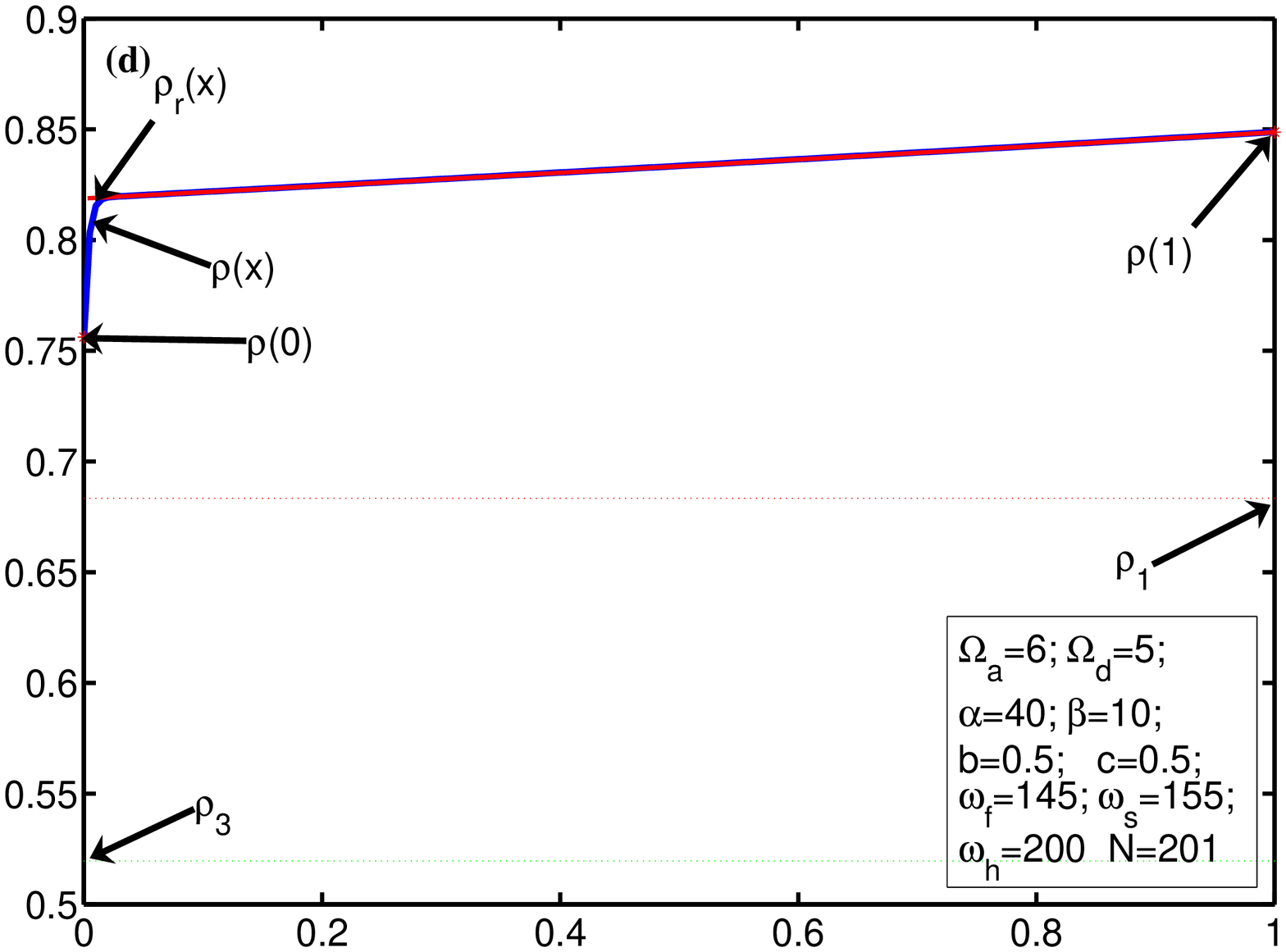}
  \caption{There are boundary layers at $x=0$ if $\rho_3\le \rho(1)\le
\rho_1, \rho_1\le \rho(0) \le 1$ {\bf (a)}; Or $\rho_3\le \rho(0), \rho(1)\le 1$ and $\rho(0)>\rho_r(0)$ {\bf (b)}; Or $\rho_3\le \rho(0)\le \rho_1, \rho_1\le \rho(1)\le 1$ {\bf (c)}; Or $\rho_3\le \rho(0), \rho(1)\le 1$ and   $\rho(0)<\rho_r(0)$ {\bf (d)}.
}\label{Fig78}
\end{figure}

\noindent{\bf (V)} For $\rho_3\le \rho(0), \rho(1)\le 1$ and
$\rho(0)\ne \rho_r(0)$, there exists boundary layer at $x=0$. For these cases, $\rho(x)=\rho_r(x)$ for $0<x\le 1$, see
 Fig. \ref{Fig56} {\bf (d)} and  Fig. \ref{Fig78}.

\section{The properties of shock}
Finally, we discuss the properties of shock briefly.
From the discussion in Sec. III, one knows that
$\rho(0)<\rho_3$ is necessary for the existence of shock. So, at
the shock position $x_s$, $\rho_l(x_s)<\rho_r(x_s)$, see Eq. (\ref{eq23}), and the height
of the shock is [see Eq. (\ref{eq37})]
\begin{equation}\label{eq69}
\left.\begin{aligned}
\varepsilon_s=\rho_r(x_s)-\rho_l(x_s)=\frac{1-2\rho_l(x_s)+\gamma
\rho_l^2(x_s)}{1-\gamma\rho_l(x_s)}.
\end{aligned}\right.
\end{equation}
The derivative of the height $\varepsilon_s$ with respect to
$\rho_l(x_s)$ is
\begin{equation}\label{eq70}
\left.\begin{aligned} \frac{\partial \varepsilon_s}{\partial
\rho_l(x_s)}=\frac{\gamma-1}{(1-\gamma\rho_l(x_s))^2}-1<0.
\end{aligned}\right.
\end{equation}
At the same time, $\rho_l(0)=\rho(0)<\rho_3$ means $\frac{\partial
\rho_l(x_s)}{\partial x_s}>0$ [see (\ref{eq25})]. So
\begin{equation}\label{eq71}
\left.\begin{aligned} \frac{\partial \varepsilon_s}{\partial
x_s}=\frac{\partial \varepsilon_s}{\partial
\rho_l(x_s)}\frac{\partial \rho_l(x_s)}{\partial x_s}<0,
\end{aligned}\right.
\end{equation}
which means, the shock height $\varepsilon_s$ decreases with the shock position $x_s$. Therefore, we only need to give the relations between shock
position $x_s$ and the model parameters $\Omega_a, \Omega_d,
\omega_f, \omega_s, \omega_h, \omega_b, c$.

\begin{figure}
  \includegraphics[width=150pt]{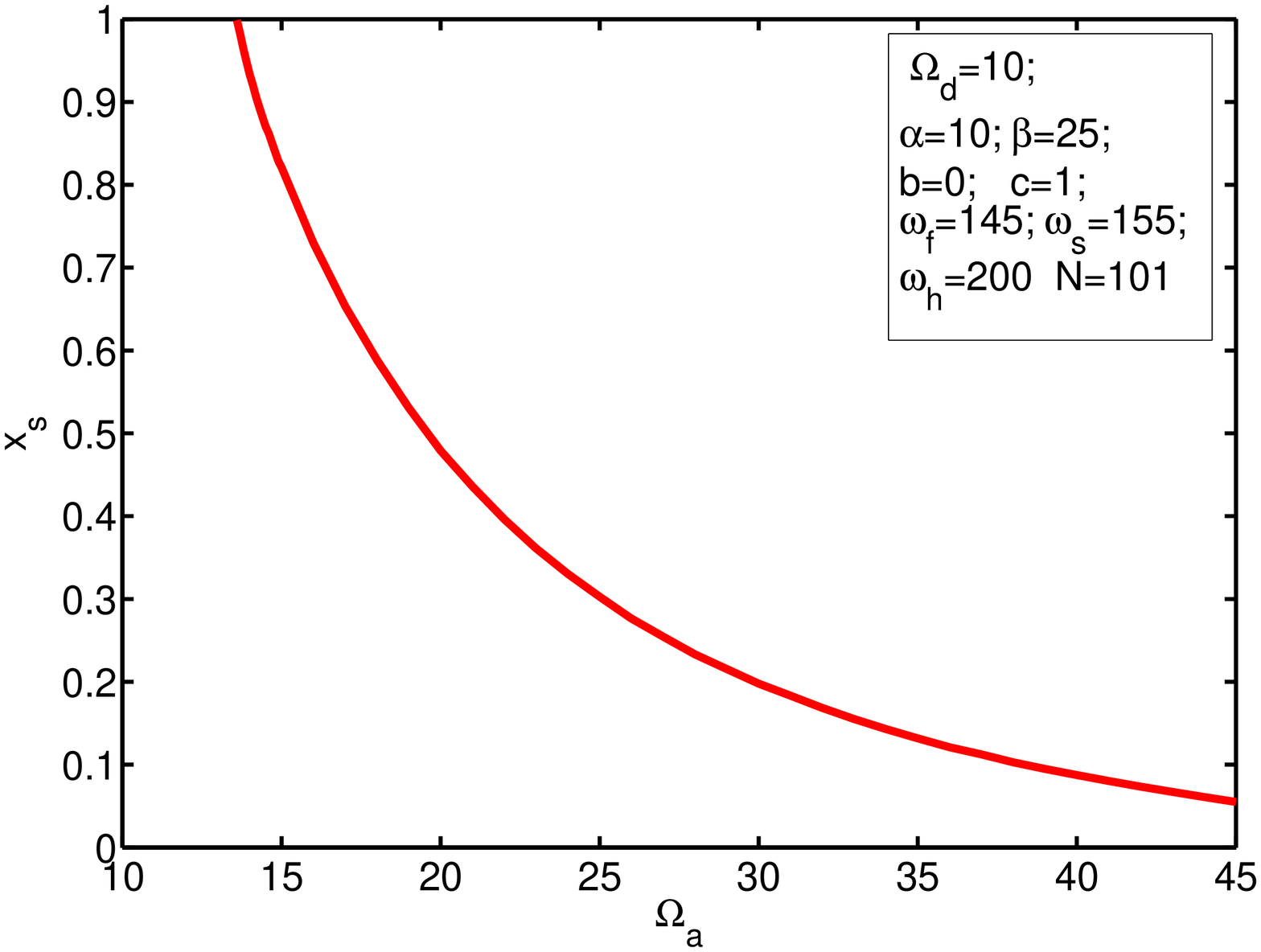}\includegraphics[width=150pt]{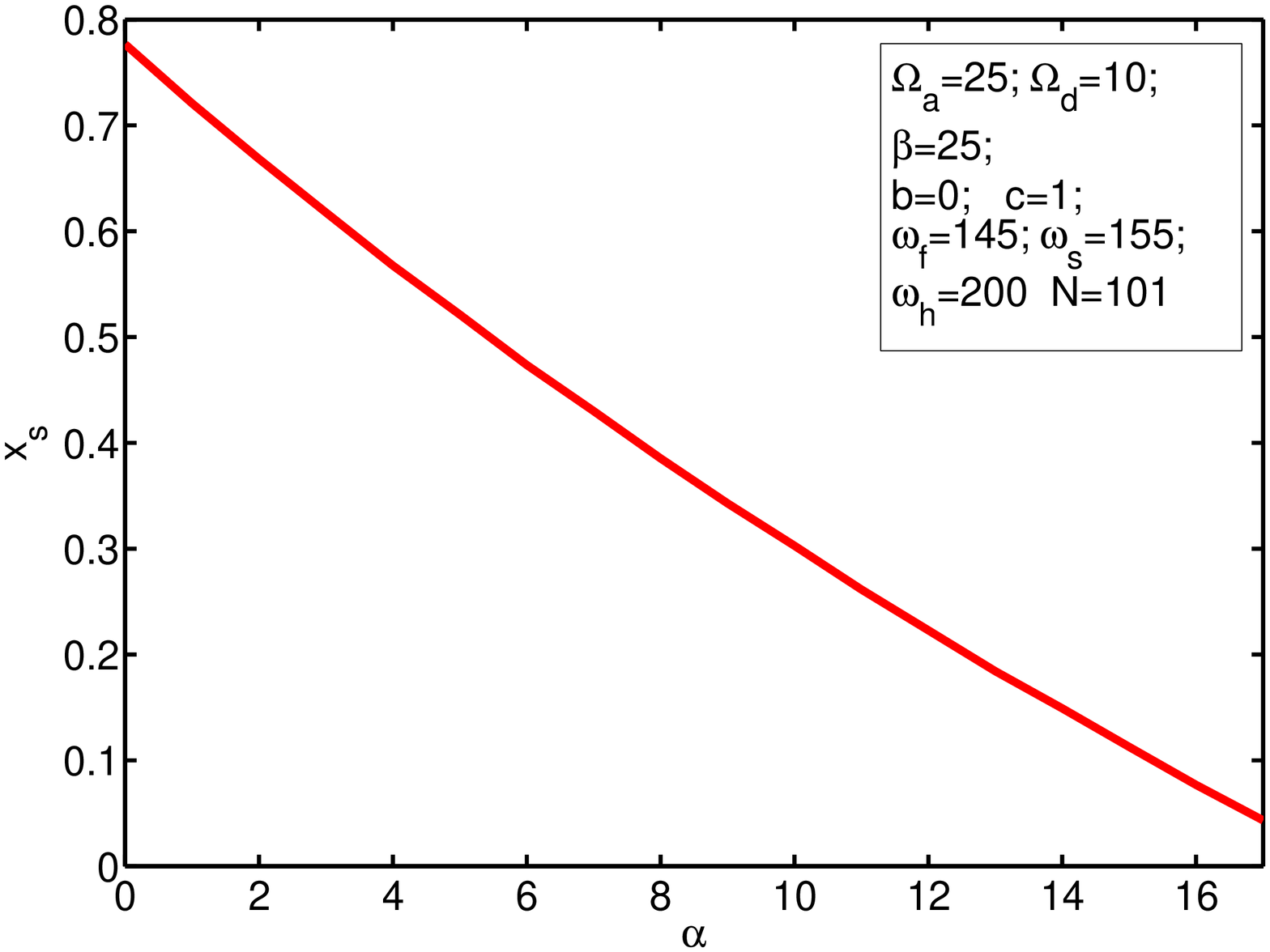}\includegraphics[width=150pt]{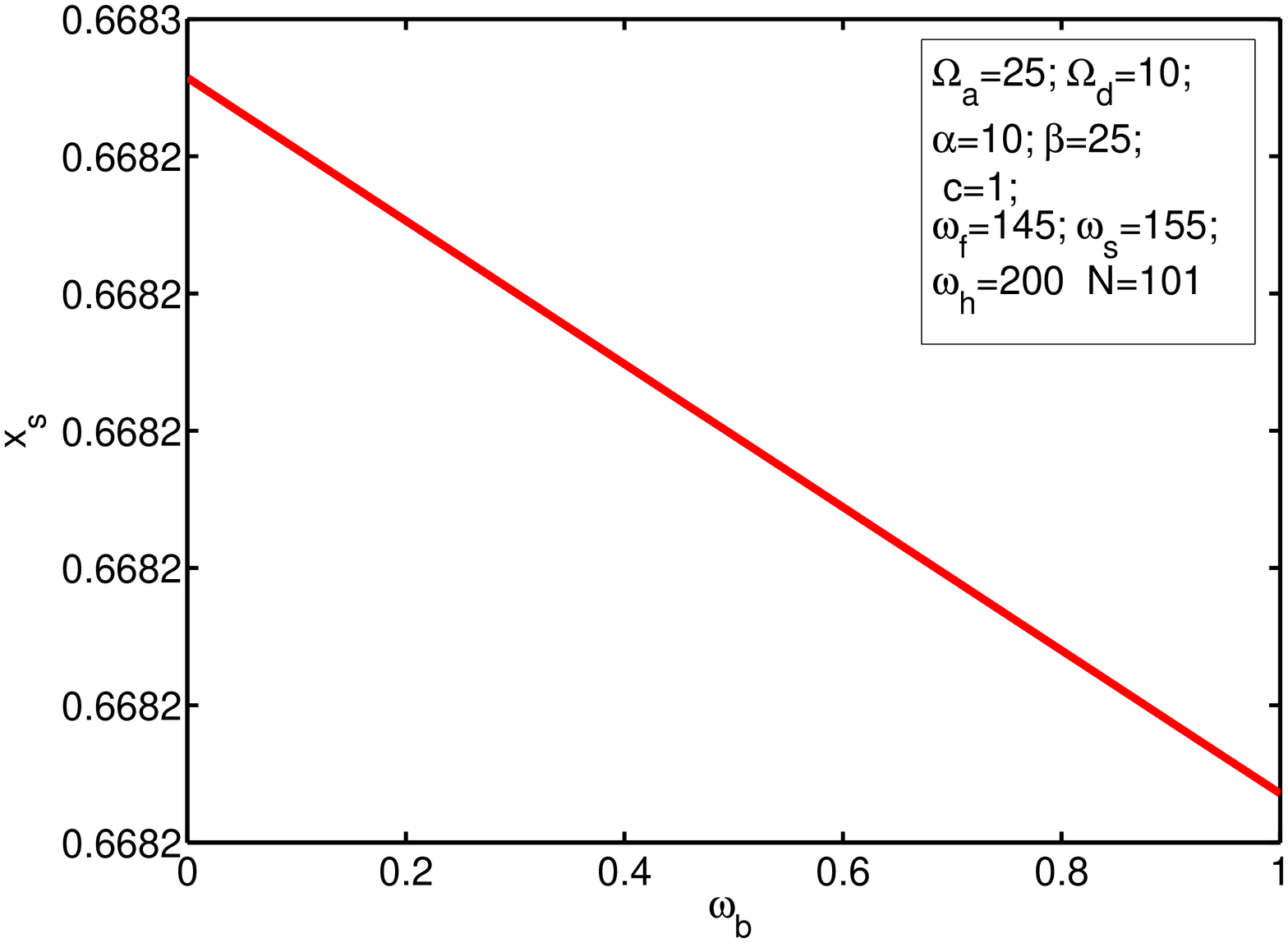}
  \includegraphics[width=150pt]{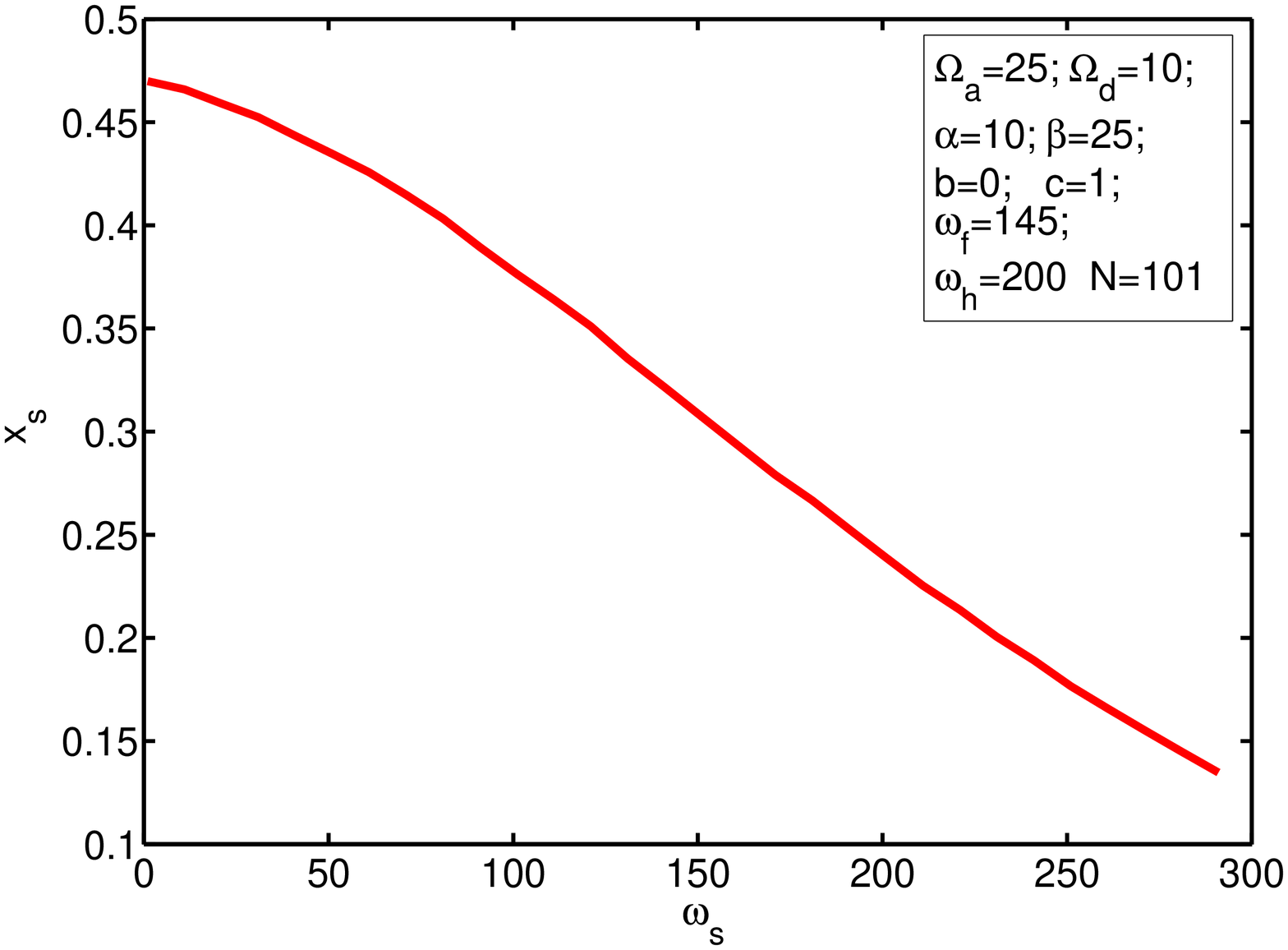}\includegraphics[width=150pt]{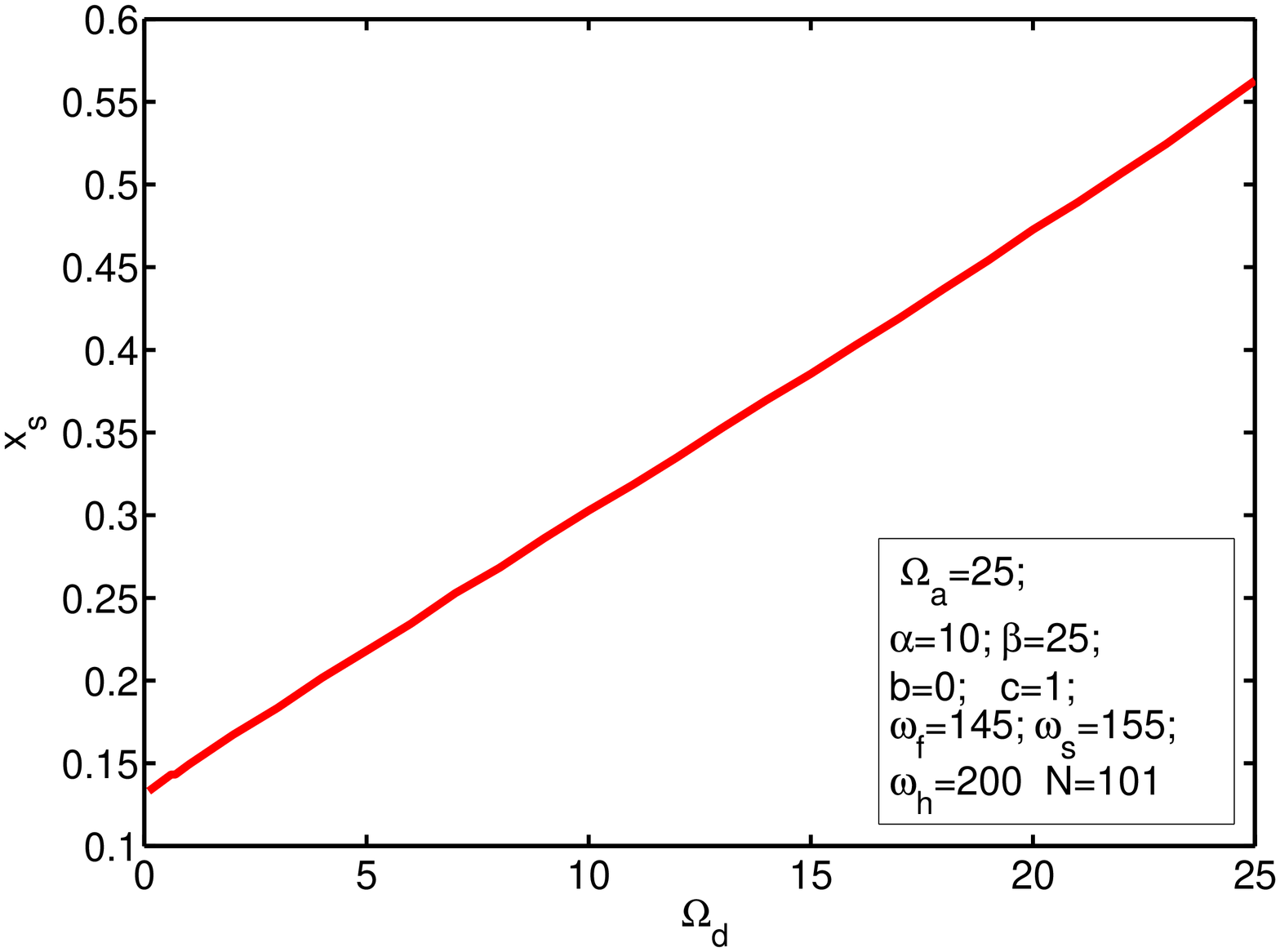}\includegraphics[width=150pt]{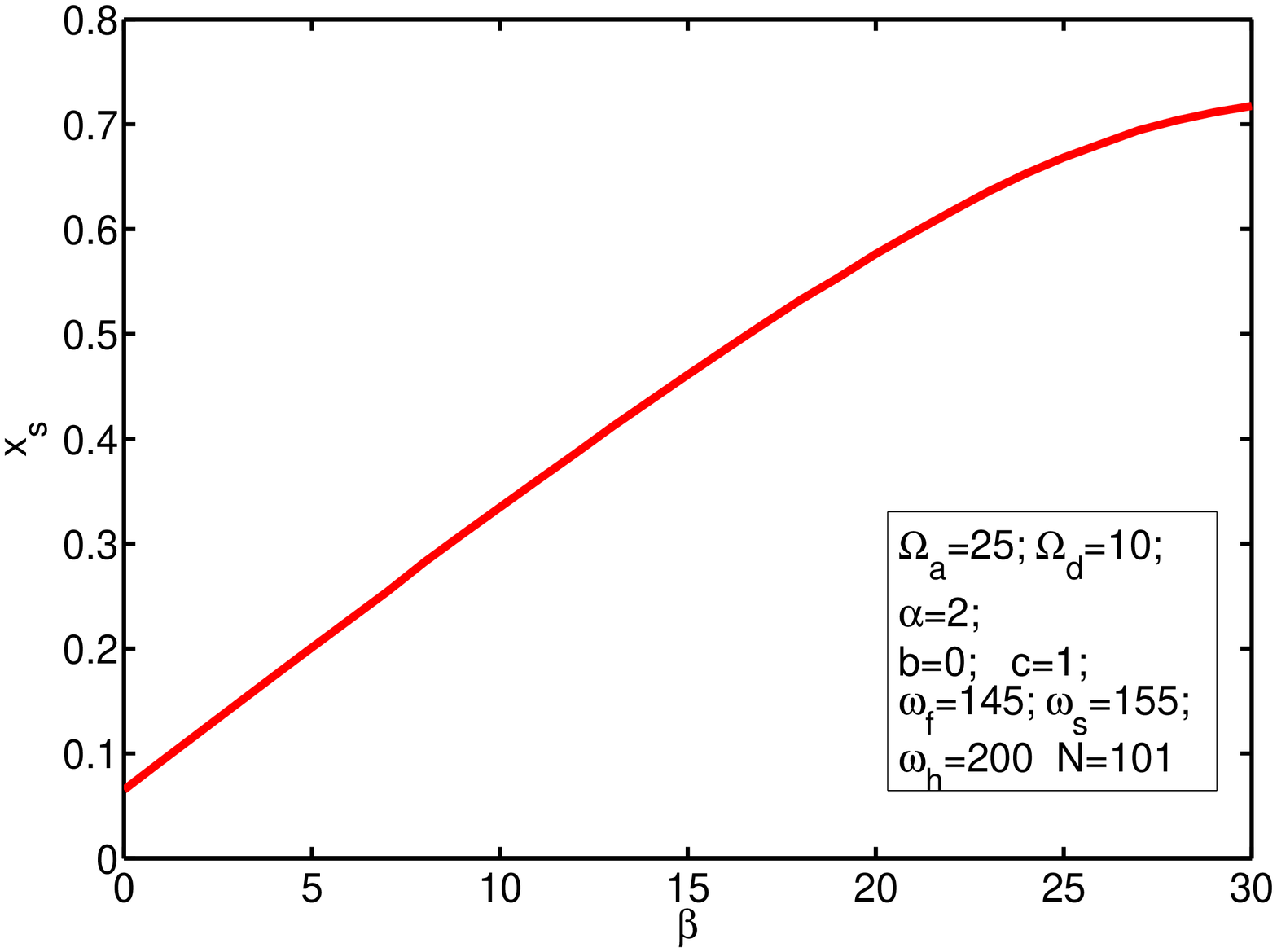}
  \includegraphics[width=150pt]{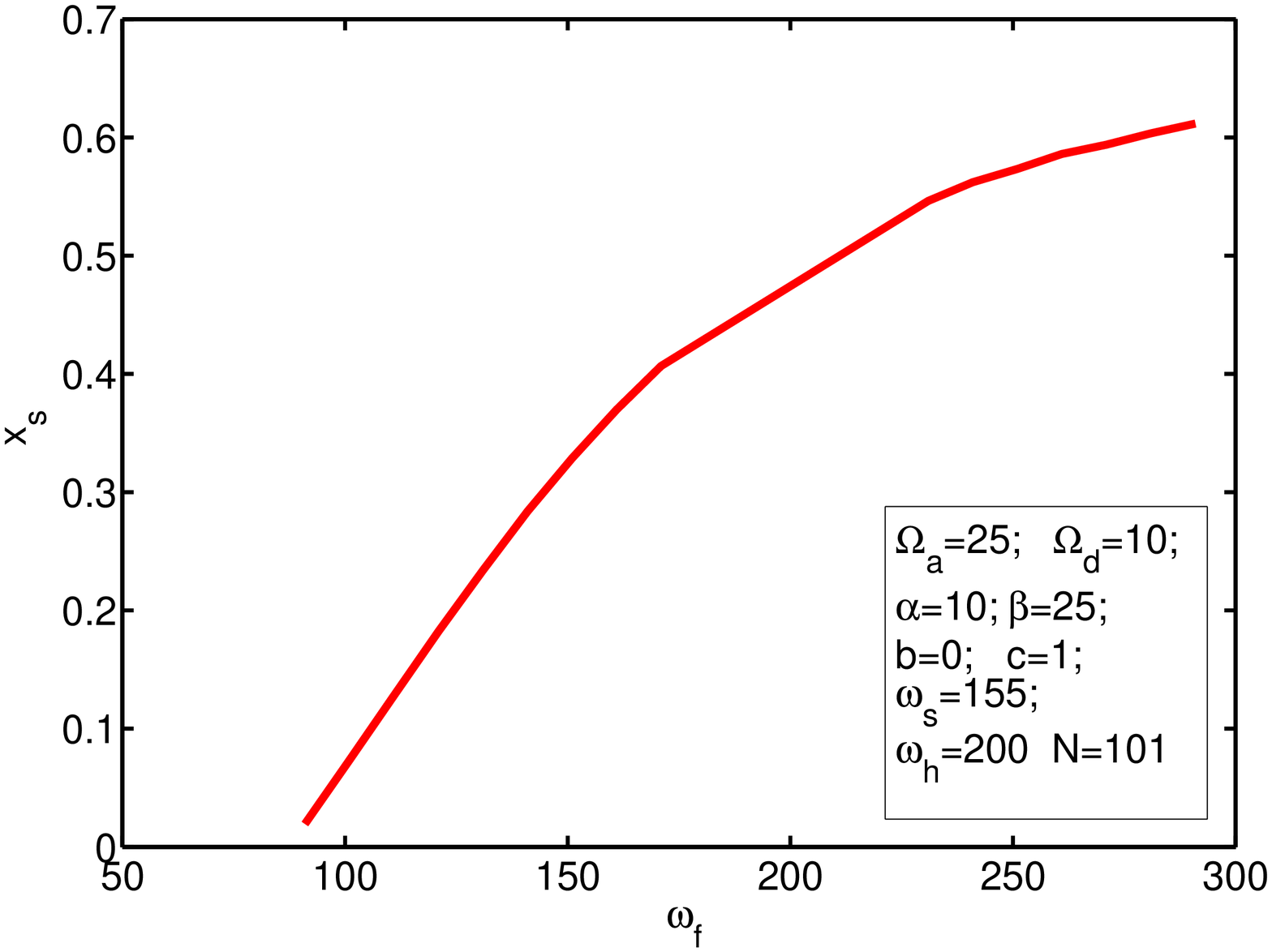}\includegraphics[width=150pt]{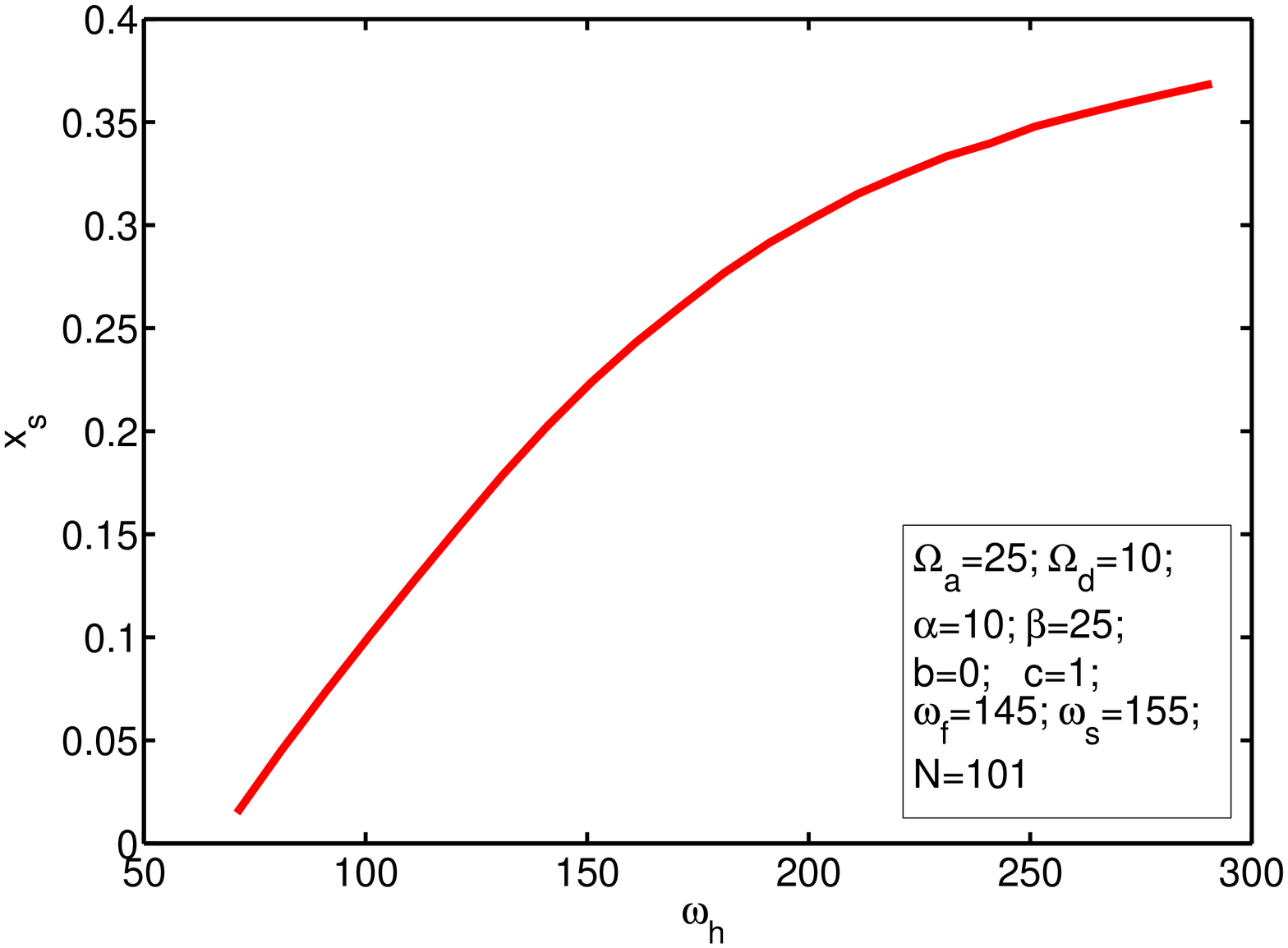}\includegraphics[width=150pt]{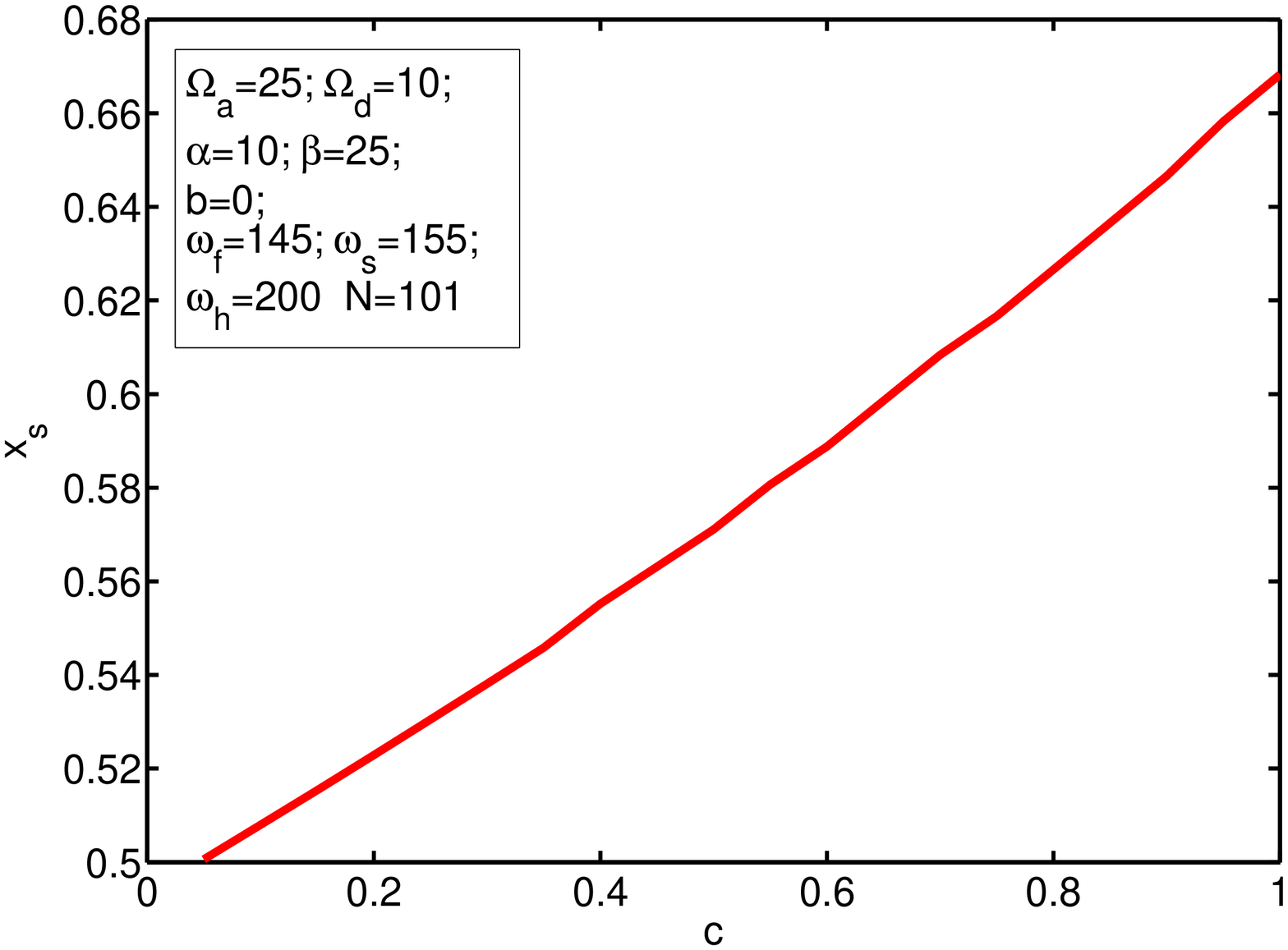}
  \caption{The shock position $x_s$ decreases with parameters $\Omega_a, \alpha, \omega_b, \omega_s$, but increases with parameters $\Omega_d, \beta, \omega_f, \omega_h, c$. In the calculations, $x_s$ is obtained by Eqs. (\ref{eq36}) and (\ref{eq33}).}\label{Fig911}
\end{figure}

Because of the complexity of the function $F$ [see (\ref{eq28})], it is difficult to get theoretical results as in \cite{Zhang2007}. From numerical calculations,
we find that, the shock position $x_s$ decreases with parameters $\Omega_a, \alpha, \omega_b, \omega_s$, but increases with parameters $\Omega_d, \beta, \omega_f, \omega_h, c$, see Fig. \ref{Fig911}. In the calculations, $x_s$ is obtained by (\ref{eq36}) and (\ref{eq33}).

\section{Concluding remarks}
In this study, the three-state process, which is presented in \cite{Nishinari2005, Nishinari2007} to model the intracellular transport of single-headed kinesin KIF1A , is theoretically analyzed using mean field approximation. By similar methods as for the usual TASEP \cite{Zhang2007}, the conditions of the existence of shock or domain wall, which is defined as the interface of low-density and high-density phases, are obtained. With the aid of numerical calculations, the parameters dependent properties of the shock are also discussed. The results obtained in this study will be helpful to understand the real biophysical properties of motor traffic in eukaryotic cells.

\vskip 0.5cm

\acknowledgments{This study is funded by the Natural
Science Foundation of Shanghai (under Grant No. 11ZR1403700).}

\newpage


\end{document}